\documentclass[useAMS,usenatbib]{mn2e}
\usepackage{graphicx,amssym}

\def\lcdm{{$\Lambda$CDM}}

\def\msun{\mbox{$M_\odot$}}

\def\mh{\mbox{$M_{\rm vir}$}}
\def\rh{\mbox{$R_{\rm vir}$}}
\def\ropt{\mbox{$R_{\rm bar}$}}
\def\ms{\mbox{$M_{*}$}}
\def\mg{\mbox{$M_{g}$}}
\def\mb{\mbox{$M_{b}$}}
\def\mstar{\mbox{$M^{\star}$}}
\def\fg{\mbox{$F_{g}$}}
\def\fgg{\mbox{$f_{g}$}}
\def\fs{\mbox{$F_{*}$}}
\def\fb{\mbox{$F_{b}$}}
\def\fst{\mbox{$F_{*}^{\rm vir}$}}
\def\fbt{\mbox{$F_{b}^{\rm vir}$}}
\def\fbU{\mbox{$F_{b,U}$}}

\newcommand{\bc}{\begin{center}}
\newcommand{\ec}{\end{center}}

\title[Mass assembly of low-mass galaxies]
      {On the mass assembly of low-mass galaxies in hydrodynamical simulations of structure formation}
\author[De Rossi et al.]
       {\parbox{17cm}{M.~E. De Rossi$^{1,2,3}$ \thanks{Email:derossi@iafe.uba.ar}, V. Avila-Reese$^4$, P.~B. Tissera$^{1,2}$, A. Gonz\'alez-Samaniego$^4$
        and S.~E. Pedrosa$^{1,2}$ }
       \\     
       \\
       $^{1}$ Consejo Nacional de Investigaciones Cient\'ificas y T\'ecnicas, CONICET, Argentina\\  
       $^{2}$ Instituto de Astronom\'ia y F\'isica del Espacio, Casilla de Correos 67, Suc. 28, 1428 Buenos Aires, Argentina\\ 
       $^{3}$ Departamento de F\'{\i}sica, Facultad de Ciencias Exactas y Naturales, Universidad de Buenos Aires, Argentina\\
       $^{4}$ Instituto de Astronom\'ia, Universidad Nacional Aut\'onoma de M\'exico, A.P- 70-264, 04350 M\'exico, D.F., M\'exico\\ }
\begin{document}

\date{Accepted  ???? ??. 2010 ???? ??}

\pagerange{\pageref{firstpage}--\pageref{lastpage}} 
\pubyear{2011}

\maketitle

\label{firstpage}

\begin{abstract}
Cosmological hydrodynamical simulations are studied in order to
analyse generic trends for the stellar, baryonic and halo mass assembly of low-mass
galaxies ($\ms \lesssim 3 \times 10^{10} \msun$) as a function of their present halo mass, 
in the context of the \lcdm\ scenario and common subgrid physics schemes. 
We obtain that smaller galaxies exhibit higher   
specific star formation rates and higher gas fractions. 
Although these trends are in rough agreement with observations, 
the absolute values of these quantities
tend to be lower than observed ones since $z \sim 2$.
The simulated galaxy stellar mass fraction 
increases with halo mass, consistently with semi-empirical inferences. 
However, the predicted correlation between them shows negligible variations up to high $z$, 
while these inferences seem to indicate some evolution. 
The hot gas mass in $z=0$ halos is higher than the central galaxy mass by a factor of 
$\sim 1-1.5$ and this factor increases up to $\sim 5-7$ at $z \sim 2$ for the smallest
galaxies.
The stellar, baryonic and halo evolutionary tracks of simulated galaxies
show that smaller galaxies tend to delay their baryonic and stellar mass assembly with
respect to the halo one.
The Supernova feedback treatment included in this model plays a key role on
this behaviour albeit the trend is still weaker than the one inferred from observations. 
At $z>2$, the overall properties of simulated galaxies are not in large disagreement with 
those derived from observations. 
\end{abstract}

\section{Introduction}
\label{sec:intro}

The hierarchical $\Lambda$ Cold Dark Matter (\lcdm) scenario offers a solid
theoretical framework for studying galaxy formation and evolution. According to this
scenario, the larger CDM virialized structures (halos)
tend to assemble systematically later than smaller ones (upsizing).
Galaxies are formed from the gas trapped into the gravitational potential
of these growing CDM structures. 
Do the assembly of baryons and stars in galaxies follow the same trend of their host
halos?
How do compare the predicted mass assembly histories with empirical inferences?

There are increasing pieces of evidence that the specific star formation 
rates
(sSFR = SFR/\ms, where \ms\ is the stellar mass)\footnote{The sSFR is a measure of the current SF activity with respect to its average past SFR, assuming that \ms\ was only generated by in-situ SF.}
of observed low-mass 
galaxies are very high up to redshifts $z\sim 1-2$. In addition, the sSFR typically increases for less massive 
galaxies \citep['downsizing in sSFR'; e.g.,][]{Bauer+2005,Feulner+2005,Salim+2007,
Noeske+2007b,Damen+2009a,Fontanot+2009,Kajisawa+2010,Karim+2011,Bauer+2011,Wuyts+2011}.  
These results suggest that, in general, small galaxies tend to delay
their stellar mass assembly presenting also a late onset of SF
\citep[see e.g.,][]{Noeske+2007b,
Bouche+2010,Firmani+2010a}. 

The relation between \ms\ and the virial mass (\mh) of the (sub)halos
have been studied widely in the local Universe by 
direct observational works 
\citep[e.g.,][and more references therein]{Mandelbaum+2006,More+2011} and also, by applying 
methods that statistically connect the observed galaxy population to that
of the \lcdm\ (sub)halos \citep[semi-empirical approach; e.g.,][and more references therein]{Vale+2004,Kravtsov+2004,
Yang+2003,Conroy+2006,Guo+2010,Rodriguez-Puebla+2013}.
According to these studies, the stellar mass fraction of galaxies ($\fs\equiv \ms/\mh$) is much smaller 
than the universal baryonic fraction ($\fbU\equiv\Omega_b/\Omega_m$) and \fs\ significantly
decreases for less massive systems. 
The baryonic fraction $\fb\equiv \mb/\mh$ 
($\mb = \ms +\mg$, where \mg\ is the cold gas mass), 
seems to follow a similar behaviour,
though the decrease is less dramatic \citep[e.g.,][]{Baldry+2008,Rodriguez-Puebla+2011, Papastergis+2012}. 

The semi-empirical approach has been extended to higher redshifts in order to obtain the evolution
of the \ms--\mh\ relation (e.g., \citealp{Conroy+2009a}; \citealp{Wang+2010}; Behroozi et al. 2010,2012; 
Moster et al. 2010,2013; \citealp{Wake+2011}; \citealp{Leauthaud+2011}; \citealp{Yang+2012}). 
By connecting the \ms--\mh\ relation obtained at different redshifts (isochrones\footnote{A relation 
between the properties of galaxies at a given epoch can be 
considered an isochrone of the evolutionary tracks of the individual systems
(e.g., the \ms--\mh\ relation).}) 
to the predicted mass aggregation histories (MAHs) of the halos, average  
evolutionary tracks for the stellar mass growth as a function of \ms\ can be derived
\citep[][]{Zheng+2007b,Conroy+2009a,Firmani+2010b,Yang+2012,Behroozi+2012,Moster+2013}. 
According to \citet[][hereafter FA10]{Firmani+2010b}, the evolutionary tracks of \ms\ corresponding to 
\ms($z=0$)$\lesssim 3\times 10^{10}$ \msun\ grow faster (at least since $z\sim 1-2$) than their 
halo MAHs. Furthermore, the difference between the evolutionary histories of \ms\ and \mh\
systematically increases for less-massive systems
('downsizing' vs 'upsizing', see e.g., Fig. 4 in FA10),
evidencing again the delay in the stellar mass assembly of smaller galaxies.
In addition, at $\ms\approx 3 \times 10^{10}$ \msun\ there seems to be
a transition from the local population of blue, star-forming galaxies to the red, quenched population
\citep[e.g.,][]{Kauffmann+2003,Weinmann+2006}. In this work, we focus the study on galaxies
with $\ms\lesssim 3\times 10^{10}$ \msun, and we will refer to them generically as "low-mass or sub-Milky Way (MW) 
galaxies". For larger (mainly red and passive) galaxies, other manifestations of downsizing
can be present \citep[see e.g.,][]{Fontanot+2009,Avila-Reese+2011a}, which
are not discussed here as they are out of the scope of the present work.

\subsection{Models and simulations of low-mass galaxies in the \lcdm\ scenario}

By using evolutionary models for isolated disc galaxies, including self-regulated SF and strong
supernova (SN)-driven galaxy outflows, \citet[][see also Dutton \& van den Bosch 2010]{Firmani+2010a} 
have shown that it is possible to reproduce the sSFRs of MW-sized galaxies 
at different $z$ as well as the local \fs-\mh\
relation. However, as the galaxy mass decreases, the sSFR systematically deviates
towards smaller values than observational inferences, and the \fs--\mh\ 
relation at higher $z$ do not agree any more with semi-empirical inferences (see also FA10). 

Regarding semi-analitic models (SAMs), it was found that small 
galaxies, both central and satellites, are too old, red, passive, and exhibit high stellar mass
fractions than what observations suggest 
\citep[e.g.,][]{Somerville+2008, Fontanot+2009, Santini+2009, Liu+2010, Guo+2011,Zehavi+2012}. 
These findings are related to the above-mentioned problem of the stellar mass buildup 
of low--mass galaxies in \lcdm\ models: these galaxies seem to assemble their \ms\
earlier than what current observations imply (see \citealp{Avila-Reese+2011a} and \citealp{Weinmann+2012},
for a discussion and more references).

With respect to cosmological N-body/hydrodynamical simulations, strong stellar-driven 
outflows at low masses are also necessary to approximate the low-mass end of the 
galaxy stellar mass function (GSMF) to the observed 
one \citep[or to attain low \fs\ values that decrease towards lower masses; for recent results, see 
e.g.,][]{Keres+2009,Oppenheimer+2010,Dave+2011,Weinmann+2012}. However, reproducing 
the GSMF and its evolution remains yet a challenge for numerical simulations, because
of resolution limitations and uncertainties in the subgrid processes \citep[but see][]{Puchwein+2012}.
In general, simulations still show a deficit of young (star-forming) low-mass galaxies at $z=0$, and
a strong decay of the cosmic SFR history since $z\sim 2$ for the low-mass galaxy population
\citep[e.g.,][]{Kobayashi+2007}.

In order to attain high resolution, the "zooming" technique of re-simulating a few
individual galaxies with higher resolution is commonly used. 
The dynamical and structural properties of low-mass "zoomed" galaxies presented in most of recent
numerical works are already in reasonable agreement with observations. 
This success seems to be partially due to the combination of a higher spatial resolution and improvements in 
the treatment of the sub-grid physics with respect to older simulations, 
in particular the inclusion of efficient SN-driven outflows. 
However, as discussed in \citet[][see also Col\'in et al. 2010]{Avila-Reese+2011b}, 
re-simulations of a few galaxies with $\mh<4\times 10^{11}$ \msun\ show that 
systems with lower masses tend to have systematically very low sSFRs and too high \fs\
with respect to observations.
Previous works that also analysed re-simulations of a few individual galaxies 
\citep{Governato+2007,Governato+2010,Piontek+2011,Sawala+2011}, 
seem to imply similar conclusions \citep[but see][who argue that 
the comparison can be improved by 
measuring galaxy properties in 
simulations by using 'artificial' observations and photometric 
techniques similar to those applied in observational works]{Brook+2012,Munshi+2013}.

Upon the understanding that the apparent problem of too early stellar mass assembly 
is generic rather than associated to a particular implementation of the current models of subgrid physics, 
it is relevant to investigate the general stellar and baryonic mass assembly of a whole population
of (mostly sub-MW) galaxies. In this way, in spite of the lower resolution, global correlations and
evolutionary trends can be obtained from a given cosmological simulation, with the
advantage of studying objects that are
not a priori selected and by implementing the same subgrid prescriptions at all scales. 

By using SPH simulations with an efficient implementation of the sub-grid physics 
\citep{Scannapieco+2006,Scannapieco+2008}, in this work we will study the stellar, baryonic, and 
dark mass assembly histories of a whole population of simulated sub-MW galaxies. 
In this way, we will be able to explore both the assembly of individual galaxies and 
the features of different observed relations as a function of redshift.
\citet{deRossi+2010} and \citet{deRossi+2012},
have already shown that the stellar and baryonic Tully-Fisher relations for these 
cosmological-box simulated galaxies agree well with observations in the local Universe.

Our aim is to discuss about the results of this numerical simulation as a {\it generic}
prediction of the assembly histories of low-mass galaxies/halos in the context of 
current models and simulations of galaxy 
evolution within the \lcdm\ cosmology, and to analyse these results in the 
light of observations.  
It is worth mentioning that new treatments for the subgrid physics 
of the model used here have been presented recently 
by \citet{Aumer+2013}.  However, as mentioned, we do not attempt to
discuss about the details of the particular implementation adopted in our work
but to analyse the general trends which are preserved and shared with other
current simulations.

The simulations are described in Section 2.
In Section 3, different relations between the properties of galaxies (e.g. sSFR--\ms, \fs--\mh, \fb--\ms\
relations) are presented at different redshifts up to $z=2$, and compared with observational
inferences.
Results at very high redshifts are analysed in Section 3.4 by using a higher-resolution run 
available only at $z\ge2$. 
In Section 4, we analyse the stellar, baryonic, and halo mass assembly histories as 
a function of present-day halo mass.
In particular, results from a parametric model of mass growth constrained to fit the empirical 
sSFR--\ms\ and \fs--\mh\ relations at different redshifts (described in the Appendix) are compared 
with the simulated trends. In Section 5, we discuss the main effects
of local and global SN-driven feedback, and analyse possible avenues to tackle the
problem of too early stellar mass assembly. Finally, our conclusions are given in Section 6.

\section{The simulations}
\label{sec:sim}

The simulations were run by using a version of the code {\small GADGET-3},
which is an updated version of {\small GADGET-2} 
optimized for massive parallel simulations of highly inhomogeneous
systems \citep{Springel+2003, Springel+2005}.
This version of  {\small GADGET-3}  includes models for 
metal-dependant radiative cooling, stochastic star formation,
chemical enrichment \citep[][]{Scannapieco+2005}, a multiphase model
for the interstellar medium (ISM) and a SN-feedback scheme \citep[][]{Scannapieco+2006}.

The chemical evolution model used in this code was developed by \citet{Mosconi+2001}
and adapted later on by \citet{Scannapieco+2005} for
{\small GADGET-2}. This model considers
the enrichment by Type II (SNII) and Type Ia (SNIa) Supernovae
following the chemical yield prescriptions of \citet{Woosley+1995}
and \citet{Thielemann+1993}, respectively. It is assumed that each SN event releases 
$0.7 \times 10^{51}$ erg, which is distributed in a fraction of $\epsilon_c=0.5$ to
the cold particles and $\epsilon_h=0.5$ to the hot particles of the multiphase 
ISM (see below).
The time-delay for the ejection of material in SNIa is randomly selected within
$[0.1,1]$ Gyr. For SNII, we assume a life-time of $\approx 10^6 $yr.

The multiphase model improves the description of the ISM as it 
allows the coexistence of diffuse and dense gas phases \citep{Scannapieco+2006,Scannapieco+2008}.
In this model, each gas particle defines its cold and hot phases by applying local entropy criteria, which 
will allow the particle to decouple hydrodynamically from particular low entropy ones if they are not part 
of a shock front.  The SN feedback and multiphase models work together at the time a cold gas particle, 
which can build up a SN energy reservoir, decides when to thermalise this energy into the ISM. Again
the decisions are made on particle-particle basis and following physically motivated criteria as explained 
in detail by \citet{Scannapieco+2006}. This allows the released SN thermal energy to play a role 
in the local properties of the ISM as well as  to drive  hydrodynamic large-scale movements (outflows).  
Furthermore, our SN feedback scheme does not include parameters that would depend on the global 
properties of the given galaxy (e.g., the total mass, size, etc.) thus making it suitable for cosmological 
simulations where systems with different masses have formed in a complex way.

In this work, we analysed a whole population of galaxies taken from a cubic box of a comoving 
14.3 Mpc side length, representing a typical field region of a $\Lambda$CDM Universe with
$\Omega_m =0.3, \Omega_\Lambda =0.7, \Omega_{b} =0.045$, a normalisation of the power 
spectrum of ${\sigma}_{8} = 0.9$ and $H_{0} =100 \, h$ km s$^{-1} {\rm Mpc}^{-1}$ with  $h=0.7$.
The simulation was run using $2 \times 230^3$ particles 
(s230), leading to a mass resolution of $8.4\times 10^{6} \ M_{\odot}$ and $1.3\times 10^{6} \  M_{\odot}$ 
for the dark matter and (initial) gas components, respectively. To check numerical effects
and to follow the galaxy assembly at high redshifts with reasonable resolution,
we use a simulation with $2 \times 320^3$ particles (s320), corresponding to a mass resolution of 
$3.1\times 10^{6} \ M_{\odot}$ and $4.9\times 10^{5} \  M_{\odot}$
for the dark matter and (initial) gas particles, respectively.
Due to high computational costs, s320 was stopped at $z \approx 2$.
These simulations are part of the Fenix project which aims at studying the chemo-dynamical 
evolution of galaxies (Tissera et al., in prep.).

We identified virialized structures by employing a standard friends-of-friends technique, while
the substructures were then individualized by applying the {\small SUBFIND} algorithm of
\citet{Springel+2001}. We constructed our main sample by using only the central galaxy in each dark 
matter halo.
However, when necessary, we also analysed the trends for the subsample of satellites.       
In order to diminish resolution issues, for s230, we restricted our study to systems
with $M_{\rm vir}\gtrsim10^{10.3} M_{\odot}$, corresponding to
a total number of particles of $N_{\rm sub} > 2000$ in the central galaxies.  
In the case of satellites, we also considered systems with $N_{\rm sub} > 2000$.
In our simulated box, we have, for instance, 214 + 46 and 187 + 22 central/satellite  
galaxies at $z=0$ and $z=2$, respectively, obeying this criterion. 
However, to construct the MAHs of galaxies, we considered all the progenitors 
identified by the {\small SUBFIND} algorithm ($N_{\rm sub} > 20$).

The main properties of galactic systems were estimated
at the baryonic radius (\ropt), defined as the one which encloses 83 per cent of the
baryonic mass of the galaxy systems. For each galaxy, we estimated its \ms,
\mg, \mb, and SFR. 
Taking into account that observational SFR tracers
are sensitive typically to $\sim 30-100$ Myr periods,
the simulated SFR is defined as the increment in stars during
a time period of 100 Myr in order to obtain
average SFR values in cases when the SF is too bursty during a given epoch.  
For each halo, we also calculated the virial radius, \rh, according 
to the spherical overdensity criteria \citep{Bryan+1998} and obtained \mh.  We also 
determined the total stellar ($M_{*}^{\rm vir}$), gas ($M_{g}^{\rm vir}$), and baryonic 
($M_{b}^{\rm vir}$) masses enclosed by \rh. 

As our simulated box corresponds to an average field
region of the Universe, there are neither clusters nor very massive galaxies.  
The most massive halos have $\mh \approx 2-3\ \times \ 10^{12}$ \msun.
This is worth noting because these simulations do not include treatments for taking into 
account the effects of active galactic nuclei (AGN) feedback, which starts to be relevant 
for halos more massive than $\mh \sim 10^{12}$ \msun\
\citep[e.g.,][]{Kobayashi+2007,Khalatyan+2008}. For the masses studied
here, the AGNs are expected to be weak or absent, both from the theoretical
and observational sides.

For more details about the simulated galaxy catalogue, see
\citet{deRossi+2010} and \citet{deRossi+2012}.  In these works,  several
structural and dynamical properties of the simulated galaxies and their correlations
are presented and discussed. These properties and correlations
are consistent with those of observed local galaxies.

\section{Analysis of the galaxy population at different redshifts}
\label{sec:results}

In this Section, we analyse several correlations between
the properties of simulated galaxies at different cosmic epochs
and compare these results with observations.

Our main analysis focus on redshifts up to $z=2$ because, at higher $z$, the 
number of galaxies resolved with $N_{\rm sub} > 2000$ is already low and 
the observations become also highly incomplete for the low masses analysed in this paper. 
Nevertheless, in \S\S \ref{sec:s320} we take advantage of the higher resolution s320
(available only at $z \ge 2$) to analyse simulated trends at very high redshifts.
s320 was ran by using the same cosmological parameters 
and settings as s230.

\subsection{sSFR as a function of mass}
\label{sec: sSFRvsMs}

\begin{figure}
\vspace{15.8cm}
\includegraphics{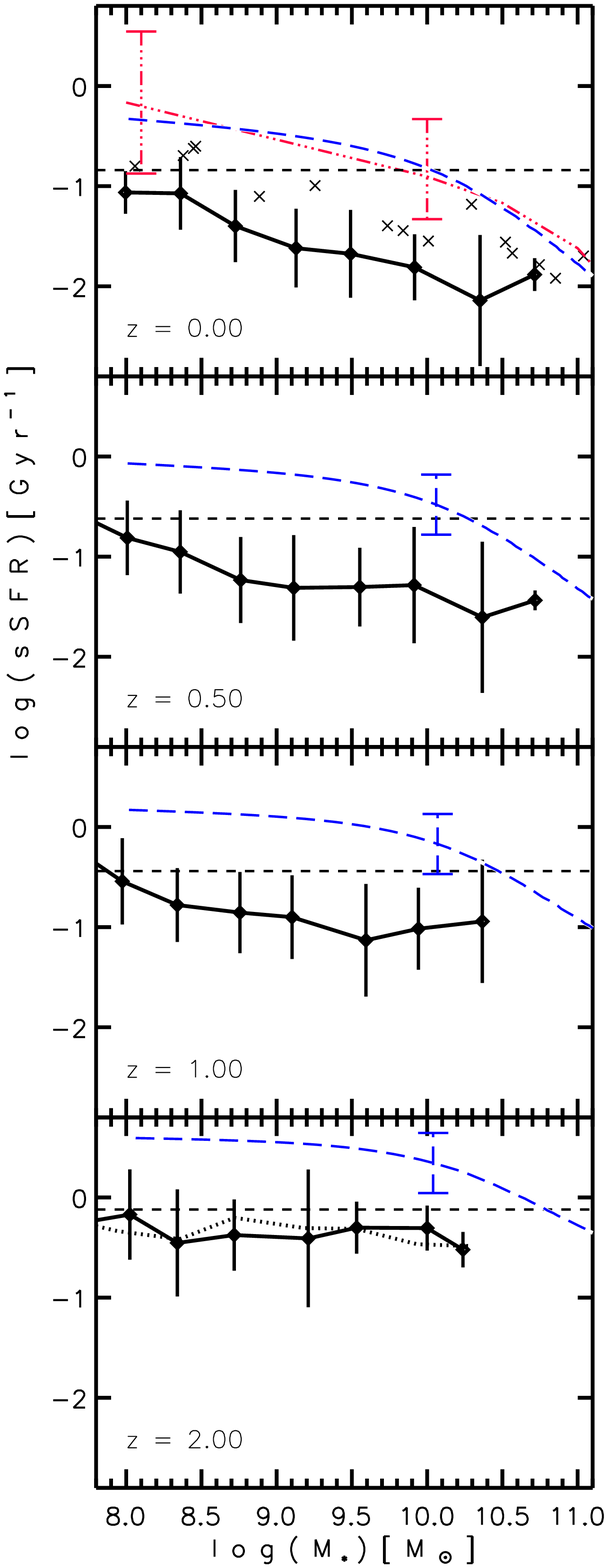}
\caption[sSFR versus $M_{*}$]
{sSFR versus $M_{*}$ for s230 at $z=0, 0.5, 1$ and $2$.  The mean relation 
and $1\sigma$ population scatter are shown with black solid lines. 
The horizontal short-dashed lines indicate the sSFR at the
given epoch corresponding to the case of constant SFR: galaxies much above (much below) 
these lines are in a current active (passive) SF phase. The blue dashed curves correspond to results
of an evolutionary toy model constrained to fit the empirical sSFR-\ms\ and \fs--\mh\ relations (see Appendix). 
At $z=0$, a fit to the observational results of \citet{Salim+2007} and its scatter is shown with 
a red triple-dot-dashed curve. 
The crosses depict a subsample of galaxies with high sSFRs in more agreement with observations 
(see Section 3.3 for a detailed discussion).
The black dotted line at $z=2$ denotes the results obtained by 
using the s320 simulation.
}
\label{fig:sSFRvsM}
\end{figure}

\begin{figure*}
\begin{center}
\resizebox{17.5cm}{!}{\includegraphics{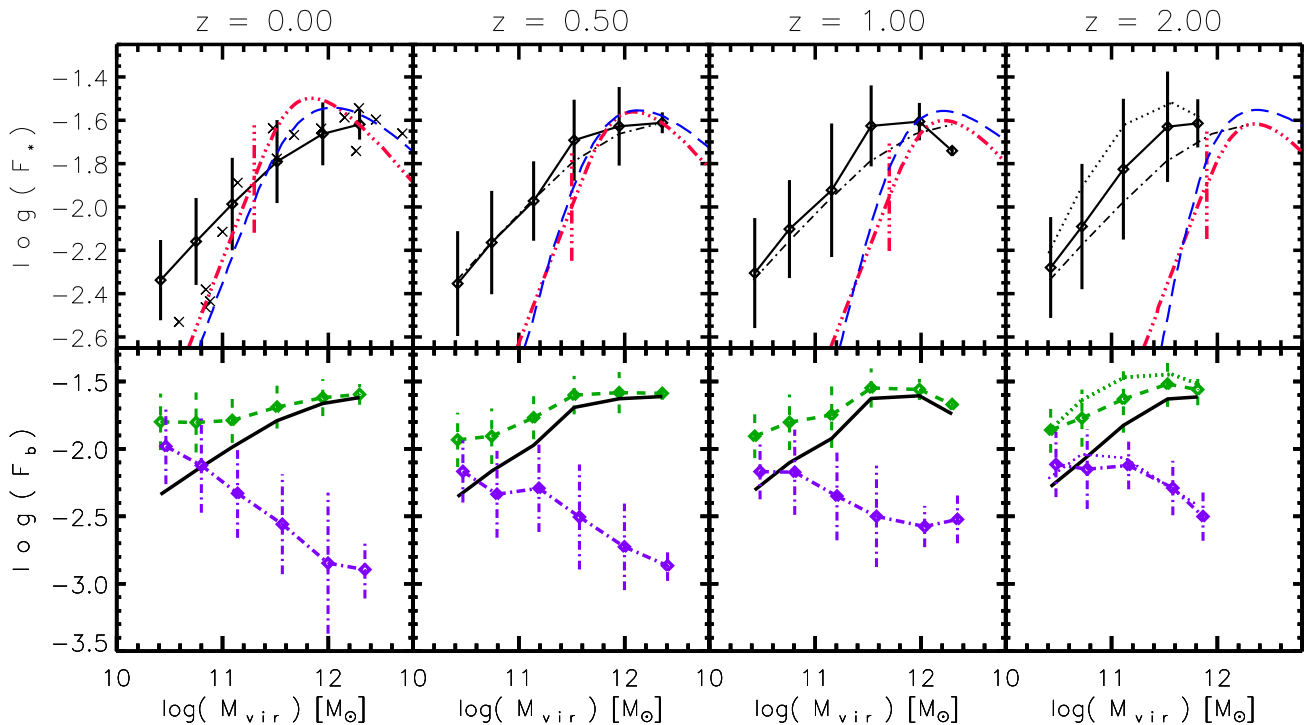}}\\
\end{center}
\caption[$f_{\rm s}$ versus $M_{\rm h}$]
{Upper panels: \fs\ versus \mh\ for s230.
The mean relation and $1\sigma$ population scatter are shown at $z=0, 0.5, 1$ and $2$ (black solid lines).
Results at $z=0$ are reproduced at higher $z$ for comparison (dot-dashed lines).
The red triple-dot-dashed curves represent         
semi-empirical inferences and their $1\sigma$ 
uncertainty, while the blue dashed curves (shown also in Fig. 1) correspond to the toy model constrained 
to fit the empirical sSFR-\ms\ and \fs--\mh\ relations (see Appendix).
The crosses depict the same high-sSFR galaxies showed in Fig. 1.
Bottom panels: Mean \fb--\mh\ (green dashed lines) and \fg--\mh\ (violet dot-dashed lines) relations 
together with the corresponding standard
deviations.
For comparison, the \fs--\mh\ curves are reproduced also in 
bottom panels (black solid curves). 
At $z=2$, the dotted lines in each panel represent the results obtained 
for \fs\ (black, upper panel), \fb\ (green, lower panel) and \fg\ (violet, lower panel) by 
using the high resolution s320 simulation.
}
\label{fig:fivsMh}
\end{figure*}

Figure \ref{fig:sSFRvsM} shows the sSFR of simulated galaxies averaged in bins 
of log\ms\ versus the averaged log\ms\ value of each bin. Vertical lines depict the 
standard deviation around the mean in each mass bin for the whole 
sample.
The horizontal short-dashed line in each panel 
shows the sSFR that a galaxy would have if it had formed its stellar component at a constant SFR; in this
case sSFR $\equiv$ SFR/\ms = 1/[(1 -- R)($t_H$($z$) -- 1 Gyr)], where $R=0.45$ is the 
assumed average gas return factor due to stellar mass loss, $t_H$ is the cosmic time, and 
1 Gyr is subtracted to take into account the onset of galaxy formation. 

In Fig. \ref{fig:sSFRvsM}, the following trends can be appreciated:
(a) simulated galaxies exhibit a tight relation between sSFR and \ms\ 
at all epochs  (the standard deviation is $\approx 0.3$ dex at $z=0$ and 
it does not increase significantly at higher $z$ for any of the mass bins);
(b) the sSFR tends to decrease with \ms\ in such a way that the relation
becomes flatter at higher $z$ (by performing a linear fit 
in the log-log plane, the slope obtained for the relation is $\approx -0.40, -0.28, -0.20,$ and $-0.06$ 
at $z=0$, 0.5, 1.0 and 2.0, respectively); 
(c) the mean sSFRs of simulated galaxies are lower than those corresponding to
the case of constant SFR since $z\sim 2$ (horizontal short-dashed lines; 
for $z>2$, galaxies tend to have a sSFR close to the case of constant SFR) being this
difference larger for more-massive systems as $z$ decreases; 
(d) the average overall sSFR of the simulated galaxy population decreases significantly 
as $z$ decreases, roughly by a factor of $\sim 10-15$ from $z=2$ to $z=0$;
the cosmic SFR in the entire box decreases by factor of $\sim 8$ during the same period.

We separate spheroid-dominated from disc-dominated galaxies 
by defining the latter as those systems with more than 75\% of 
their gas component on a rotationally supported disc structure
by using the condition $\sigma / V < 1$ to select them.
For more details, see \citet{deRossi+2010} and \citet{deRossi+2012}. 
All the other systems are considered to be spheroid-dominated.
We did not find a significant difference between the mean sSFR-\ms\ 
relations of both groups.
We have also separated central galaxies from satellites, and found that satellites exhibit
a wider distribution of sSFRs than central systems. If any, the former systems have
slightly larger sSFRs than the latter at a given \ms\ for $z\lesssim 1$. The fraction 
of satellite galaxies in our simulation is actually small (0.21, 0.14, 0.13, and 0.12 at 
$z=0.0$, 0.5, 1.0 and 2.0, respectively), so that they hardly contribute 
to the mean sSFR-\ms\ relation. The  number of small satellites is determined by numerical resolution.

The blue dashed curves in Fig. \ref{fig:sSFRvsM} denote fittings to observations  
derived from a parametric toy model
(for a compilation of observations and their comparison with these fits, see Fig. \ref{Mod-sSFR} in the
Appendix). Simulated galaxies exhibit lower mean sSFRs than those derived from observations, being the
difference larger as $z$ decreases \footnote{Because observers measure the SFR 
with different surface brightness limits, the comparison with our simulations could be sensitive to the way
in which we define the radius of galaxies. We have calculated the SFR and \ms\ of simulated systems at two
other radii: 0.5 and 1.5 $R_{\rm bar}$. In both cases, we verified that we obtained sSFR--\ms\ relations 
that are very close to the one
shown in Fig. \ref{fig:sSFRvsM} for quantities measured at $R_{\rm bar}$. Therefore, the uncertainty
about the radius at which the SFR and \ms\ are measured observationally does not seem to play a significant
role in the comparison with simulations.}. 
Moreover, most of simulated galaxies present sSFRs below the line associated to a constant SFR 
(black short-dashed lines).  The deviations from the case of constant SFR 
increase with stellar mass and decrease with redshift,
suggesting that the active phases of SF in simulated galaxies took place early, $z>1$. 
Nevertheless, as noted before,
the simulations seem to be able to reproduce qualitatively the observed increase of
sSFR for less-massive systems at low redshifts (downsizing in sSFR) and are also able to predict 
the flattening of the sSFR--\ms\ correlation at higher redshifts. 
We will see that our SF and SN-feedback scheme in a multi-phase ISM plays a crucial role
on generating these trends.

\subsection{Stellar and baryonic mass fractions} 
\label{sec: fivsm}

In Figure \ref{fig:fivsMh}, we can appreciate the
dependences of \fs\ (upper panels), \fb\ and a similar defined fraction for the 
gas-phase, $\fg\equiv \mg/\mh$ (bottom panels) on \mh. 
Results are shown for the same four redshifts represented in Fig. \ref{fig:sSFRvsM}.

The main trends obtained for \fs\ (upper panels)
can be summarised as follows: 
(a) \fs\ systematically increases with \mh\
at all available $z$, though, for our most massive galaxies, evidence of reaching
a maximum at $\mh\sim 10^{12}$ \msun\ is seen; 
(b) even at this maximum, the mean values of \fs\ are not larger than $0.025$;
(c) at $\mh\lesssim 10^{11}$ \msun, the \fs--\mh\ correlation does not evolve significantly
since $z \sim 2$ (roughly $\fs\propto M_{\rm vir}^{0.45}$) 
while, at $\mh\gtrsim 10^{11}$ \msun,
it becomes slightly steeper as $z$ increases;
and (d) the scatter around the \fs--\mh\ correlation is 
$\approx 0.2-0.3$ dex at $z=0$, increasing towards higher $z$.

The red triple-dot-dashed curves in the upper panels show the average \fs--\mh\
correlations inferred by matching the observed GSMFs
to the halo/subhalo mass functions\footnote{In this approach, the mass function of pure 
dark matter halos is used but the masses of halos (dark+baryonic matter) could end 
up smaller when baryonic processes are considered (e.g., as a result of ejections of gas out of 
the virial radius in low-mass halos). \citet{Avila-Reese+2011b} reported that, because of
gas loss, low-mass halos can be $\sim 15\%$ less massive than in pure dark matter simulation,
implying a slightly higher \ms-to-\mh\ ratio at low masses than those shown in Fig. \ref{fig:fivsMh}
(see their Fig. 6). Recently, \citet{Munshi+2013} have reported that this fraction could be 
even higher, up to $\sim 30\%$.}.  
As described in FA10, results were obtained by parametrising a continuous function according to
the data reported in \citet{Behroozi+2010} 
in the separate redshift ranges $0\lesssim z \lesssim 1$ and $1\lesssim z \lesssim 4$. 
The error bar in each panel depicts an estimate 
of the uncertainties, which are mainly dominated by the systematic uncertainty in the determination 
of \ms\ \citep{Behroozi+2010}.  
We can appreciate that the slope of the semi-empirical \fs--\mh\ relation tends 
to be steeper than the simulated one since $z \sim 2$.  
In particular, simulated results at $z=0$  are close to those associated
to the semi-empirical constraints at $\mh \gtrsim 10^{11}$ \msun, deviating
towards greater \fs\ at the low-mass end of the relation.
Note, however, that the latter differences should be smaller if the semi-empirical approach
used a halo mass function derived from simulations that include baryons (see the footnote).
Regarding the evolution with redshift, the semi-empirically inferred relation shifts systematically towards 
larger \mh\ as $z$ increases, while the simulated relation exhibit negligible variations since $z \sim 2$.
We do not find significant 
differences between the \fs--\mh\ correlations associated to disc-dominated and spheroid-dominated systems.

In the bottom panels of Fig. \ref{fig:fivsMh}, we analyse the evolution
of \fb\ and \fg.  As \mh\ decreases,
\fb\ significantly increases with respect to 
\fs, indicating the presence of higher gas fractions in the case
of less-massive galaxies.  As a result, the predicted \fb--\mh\ correlations are much flatter than the
\fs--\mh\ ones, specially at lower $z$ (in the bottom panels, the 
\fs--\mh\ correlations are reproduced again with black solid curves for comparison). 
\fb\ reaches the maximum values in the case of
most massive halos ($\sim 2\times 10^{12}$ \msun), with $\fb\approx 0.025-0.030$.
The latter values are 
much smaller than the universal baryon fraction (\fbU=0.15, for the cosmology adopted here),
suggesting that significant outflow events may have affected these systems \citep[see also][]{deRossi+2010}.
Note that we consider the whole (cold + hot) gas component in our analysis. 
Nevertheless, for most simulated galaxies, 
more than 90\% of the gas mass inside $\ropt$ is
cold ($T<15000$ K).  The fraction of gas in the hot phase does not attain more than $\sim 20\%$ of \fg\
for any of the systems,
with the greater values associated only to most massive galaxies at $z=0$.

Regarding the evolution of the \fg--\mh\ 
correlation (violet dotted-dashed lines), we can appreciate that since $z\sim 2$: 
for halos with $\mh\gtrsim 5\times 10^{10}$ \msun, \fg\ decreases with time,
while for less massive halos, it systematically increases. 
At $\mh\sim 5\times 10^{10}$ \msun, 
$<\fg>\approx <\fs>\approx 0.008$ (i.e. $\mg/\ms\sim 1$).  
The increase of \fg\ in the case of low-mass galaxies 
suggests that they have lower star formation efficiencies.

\begin{figure}\hspace*{0.5cm}\\
\vspace{7cm}
\includegraphics{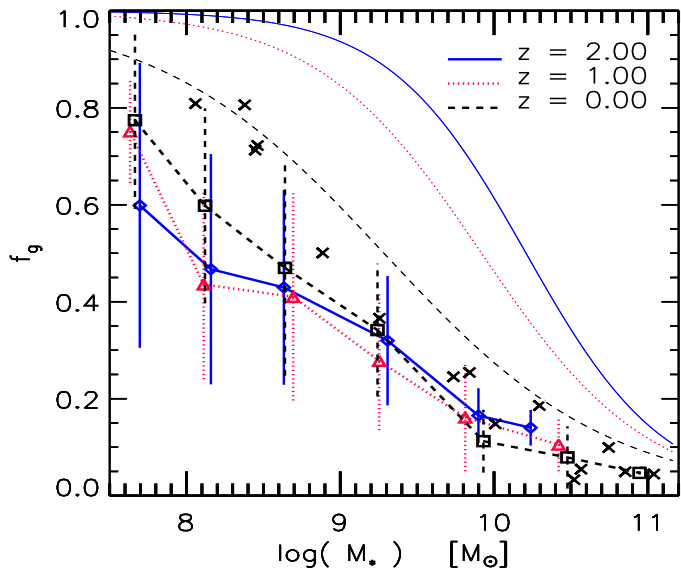}
\caption{
Average galaxy gas fraction vs \ms\ at $z=2, 1, 0$ (blue solid, red dotted and black dashed lines, respectively). 
The error bars depict the $1\sigma$ population scatter. The dotted curves without error bars correspond to analytical fits to observations given
in \citet{Stewart+2009}. Crosses denote results for the same galaxies with high-sSFR shown in Fig. \ref{fig:sSFRvsM}.
}
\label{fig:fgasvsz}
\end{figure}

Finally, in Fig. \ref{fig:fgasvsz}, we analyse the galaxy gas mass fraction ($\fgg\equiv \mg/ \mb$) 
as a function of \ms\ at $z=0, 1$, and 2.  Though the scatter is large, the simulations are able 
to reproduce at all epochs the observed trend of decreasing \fgg\ with
\ms. The relation becomes slightly steeper as $z$ decreases in such a way
that less-massive galaxies exhibit higher \fgg\ towards $z \sim 0$. 
Fig. \ref{fig:fgasvsz} also shows results from a generalized analytical fit to 
the observed cold \fgg\--\ms\ relation \citep{Stewart+2009}. We see that, at a given
\ms, observed (cold) gas fractions
tend to be higher than those obtained for simulated galaxies, where most of the gas is actually in the cold phase. This is 
consistent with the fact that the sSFRs of the simulated systems are lower than observed 
ones (Fig. \ref{fig:sSFRvsM}): simulated galaxies seem to have less cold gas available to fuel SF and 
more significant stellar mass fractions than observed galaxies.

\subsection{Simulated galaxies in agreement with observations}

In Fig. \ref{fig:sSFRvsM}, we present the $\sim 7\%$ of galaxies with the highest sSFRs at $z=0$,
which have gas fractions similar to those reported by observers. All these galaxies have sSFRs 
within the $1\sigma$ scatter associated to observations along almost three orders of 
magnitude in \ms, though slightly below the average observed values. 
The \fs--\mh\ and \fgg--\ms\ relations for these galaxies at $z=0$ are
also in general agreement with observations
(Figs. \ref{fig:fivsMh} and \ref{fig:fgasvsz}, respectively). 
Therefore, our simulations were able to produce at least 
15 galaxies that are within the scatters of the observational correlations studied here, 
while actually the average relations of the whole population significantly deviate from observed ones. 
Taken into account that these 15 systems are above the $1.5\sigma$ of the simulated distribution, 
these findings shows the relevance of studying the behaviour of the whole
population in order to draw generic conclusions about average correlations of galaxies.
Nevertheless, it is encouraging
that simulations can generate some systems which have similar properties to observed galaxies in the Local universe.

\begin{figure}
\vspace{11.8cm}
\includegraphics{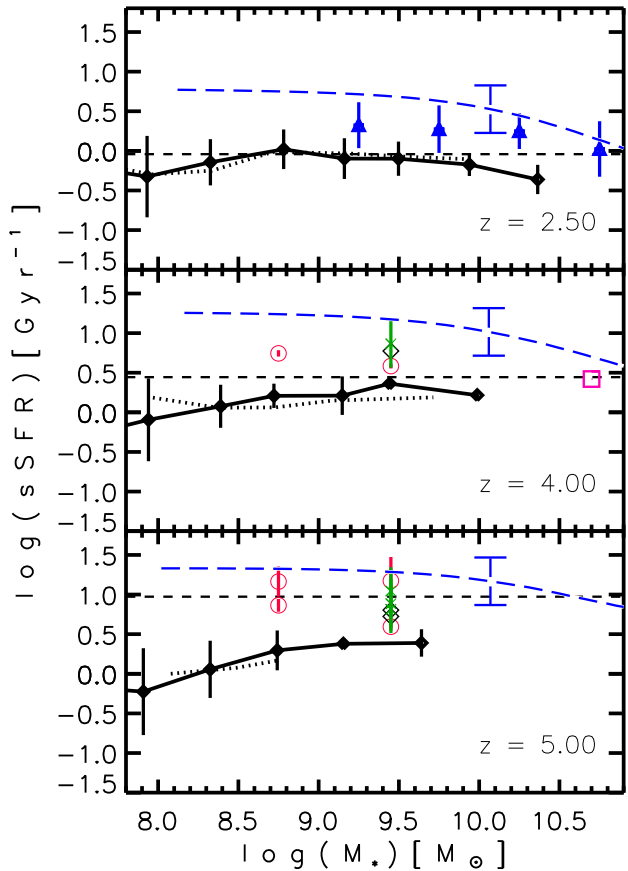}
\caption[sSFR versus $M_{*}$ at $z>2$]
{
sSFR versus $M_{*}$ for the high-resolution s320 simulation at $z=2.5, 4.0$ and $5$.  
See Fig. \ref{fig:sSFRvsM} for results at lower redshifts. 
The mean relation
and $1\sigma$ population scatter are shown with black solid lines.  
The dotted lines show results in the case of s230 for comparison. 
The horizontal short-dashed lines indicate the sSFR at the
given epoch corresponding to the case of constant SFR.
Results from different observational works are shown:
\citet[][blue triangles]{Bauer+2011}, 
\citet[][green crosses]{Stark+2013}, 
\citet[][red circles]{Gonzalez+2012},
\citet[][black diamonds]{Bouwens+2012} and \citet[][pink square]{Daddi+2009}.
The last three set of data correspond to star-forming galaxies. 
The blue dashed curves correspond to results
of the evolutionary toy model (see Appendix).
}
\label{fig:sSFR-s320}
\end{figure}

\begin{figure}\hspace*{0.5cm}\\
\vspace{7cm}
\includegraphics{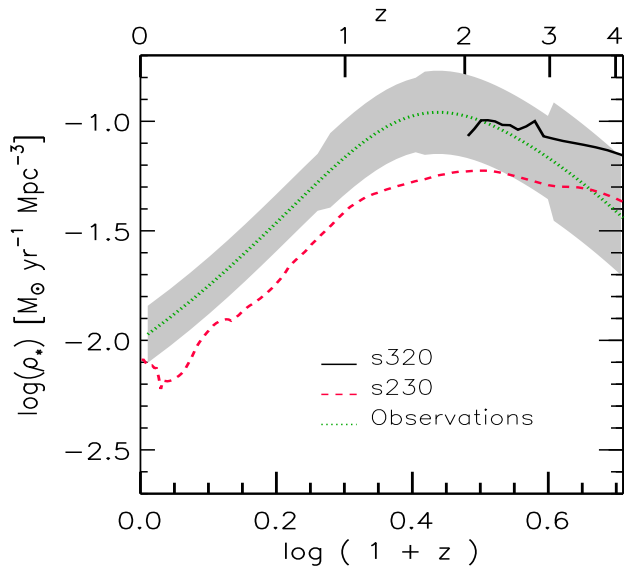}
\caption{
Cosmic SFR evolution for s230 (red dashed line) and s320 (black solid line; available only at $z\ge2$).
The green dotted line and shaded area depict the mean relation and standard deviations corresponding
to a fit to compiled observations given in \citet{Behroozi+2012}.
}
\label{fig:cosmicSFR}
\end{figure}  

\subsection{Simulated galaxies at very high redshifts}
\label{sec:s320}

In the case of s230, a significant fraction of galaxies have $N_{\rm sub} < 2000$ 
at $z>2$. Instead, for s320 (available at $z \ge 2$), galaxies are yet well resolved 
since $z\sim 5-6$. Therefore, we employed the latter simulation for exploring the overall evolutionary trends of galaxies
at very high redshifts and compared the results with the scarce available observational data (at $z>2$, most of observational studies
are complete only for $\ms> 5\times 10^{9}$ \msun). 

Fig. \ref{fig:sSFR-s320} shows the sSFR as a function of \ms\ at $z=2.5, 4.0$ and 5.0
(see Fig. \ref{fig:sSFRvsM} for comparing these findings with results at lower redshifts). The 
black solid lines with error bars denote the mean relation 
and standard deviations corresponding to simulated galaxies in s320. Results
obtained by using the s230 run are shown with dotted lines for comparison. The dashed horizontal lines indicate the case of constant 
SFR history ($R=0.30$ has been used for these early times).
Findings from different observational works are also shown:
\citet[][blue triangles]{Bauer+2011},
\citet[][green crosses]{Stark+2013},
\citet[][red circles]{Gonzalez+2012},
\citet[][black diamonds]{Bouwens+2012} and \citet[][pink square]{Daddi+2009}.
The last three set of data correspond to star-forming galaxies.
We can see that, along the mass range where simulations and observations can be compared,
the former are somewhat below the latter. 

Figure \ref{fig:cosmicSFR} shows the cosmic SFR history in our simulated box for s320 (black solid line) as
well as for s230 (red dashed line). The green dotted line and shaded area depict the mean relation and standard
deviations corresponding to a compilation 
of observations given in \citet{Behroozi+2012}. 
The peak of the 
cosmic SFR is attained at $z\sim 2$, both in s320 and s230.
Results from the high-resolution s320 run are in rough 
agreement with observations since $z \sim 5$ up to $z=2$. At higher $z$, the simulated cosmic SFR 
tends to be larger than what observations suggest, indicating that the gas is efficiently transformed into stars
at very early times.
However, the percentage of stellar mass assembled in this 
short time period is small with respect to the mass assembled at later epochs. 

\begin{figure}\hspace*{0.5cm}\\
\vspace{6.5cm}
\includegraphics{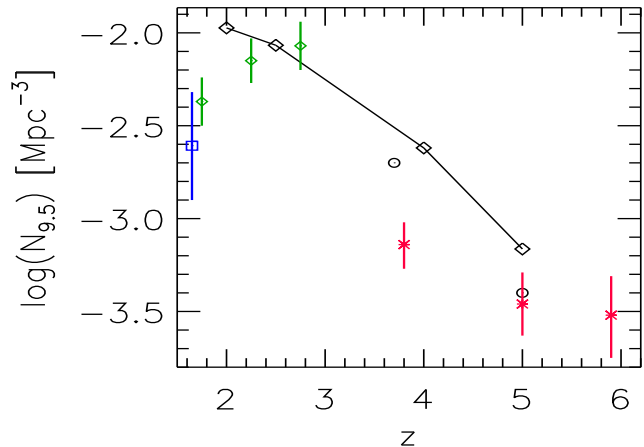}
\caption{
Number density (per unit of comoving volume) of galaxies with masses log(\ms/\msun)$=9.5 \pm 0.25$ as a function of redshift for s320 
(black solid line).  Observational measures are also shown: \citet[][blue square]{Marchesini+2009}, 
\citet[][green diamonds]{Mortlock+2011}, \citet[][red asterisks]{Gonzalez+2011}, and \citet[][black circles]{Lee+2012}. 
}
\label{fig:Nz}
\end{figure}

Finally, as the \ms--\mh\ relations are poorly observationally constrained at high $z$, we present here
the number density (per unit of comoving volume) of galaxies with log(\ms/\msun)$=9.5\pm 0.25$ ($N_{\rm 9.5}$,
black solid line in Fig. \ref{fig:Nz}) as a function of $z$. Note that these galaxies
have the largest masses in our box at these early epochs.  
A few observational studies were able to determine galaxy 
abundances down to these masses at high $z$. We plot some of such abundances in Fig. \ref{fig:Nz}, using in all 
cases $h=0.7$. 
The simulated $N_{\rm 9.5}$ follows the same trend with z than observations, though with larger values. 
This is consistent with the 
high simulated \fs\ obtained for the s230 run at $z\sim 1-2$ with respect to the semi-empirical
determinations (Fig. \ref{fig:fivsMh}). 
The fact that the number density of the largest simulated galaxies 
is already close to that derived from observations at these high $z$ indicates
that the trend of assembling stellar mass earlier than empirical inferences is systematically
stronger for smaller systems.

After the submission of this paper, a similar analysis was performed at $z \ge 2$ by using 
a cosmological hydrodynamical simulation in a box of $114^3$ Mpc$^3$ side-lenght \citep{Kannan+2013}. 
The simulated volume used by these authors is larger than the corresponding to s320 but
the resolution is much lower.  In spite of these differences, the trend and evolution 
predicted for the sSFR--\ms\ relation, the GSMF and the cosmic SFR history are consistent
with those reported here at $2 \le z \le 5$, at least in the mass range where comparison is
possible.
We note that, at $z>2.5$, $N_{\rm 9.5}$ tends to be slightly lower in the the case of \citet{Kannan+2013}
than what the s320 run suggests (by $\sim 0.2-0.3$ dex), being these differences smaller when using the s230 run. 
However, at $1 \times 10^{11} \msun < \mh\ < 3\times 10^{11}$ \msun, their stellar mass fractions are lower than 
the ones obtained here by $\approx 0.7$ and $\approx 0.5$ dex at $z=2$ and $z=4$, respectively. 
These differences may be partially caused by
the lower GSMF and $N_{\rm 9.5}$ abundances reported by \citet{Kannan+2013}, but it is
also important to consider that the virial masses of our halos become smaller at higher $z$, 
than those measured in pure dark matter large simulations, showing a delay in their mass assembly 
(see also the upper panel of Fig. \ref{fig:massagg}).  The latter trend could be due to an 
environmental effect associated to our smaller volume. In spite of these issues, we find that the main
evolutionary features of our low-mass galaxies at $z \ge 2$ 
are similar to those reported in \citet{Kannan+2013}.

\section{The mass assembly histories of galaxies and their halos}
\label{sec: assembly}

In this Section, we analyse the evolution of individual simulated systems 
selected from s230 according to their present-day mass with the aim of exploring 
their assembly histories since $z=2$.
We focus only on those central galaxies 
with $\mh\ \gtrsim10^{10.3} M_{\odot}$ at $z=0$ 
($N_{\rm sub} > 2000$) and follow the evolution of their main
progenitors\footnote{We define the main progenitor as the one which
has the higher baryonic mass at a given time step.  See 
\citet[][]{deRossi+2012} for more details}
back in time.  In order to determine
reliable evolutionary trends, 
when reconstructing the evolutionary histories,
we consider that all the substructures identified 
in the simulated box ($N_{\rm sub} \ge 20$) are plausible progenitors.

\subsection{Galaxy vs halo mass assembly}

\begin{figure*}
\begin{center}
\resizebox{7.8cm}{!}{\includegraphics{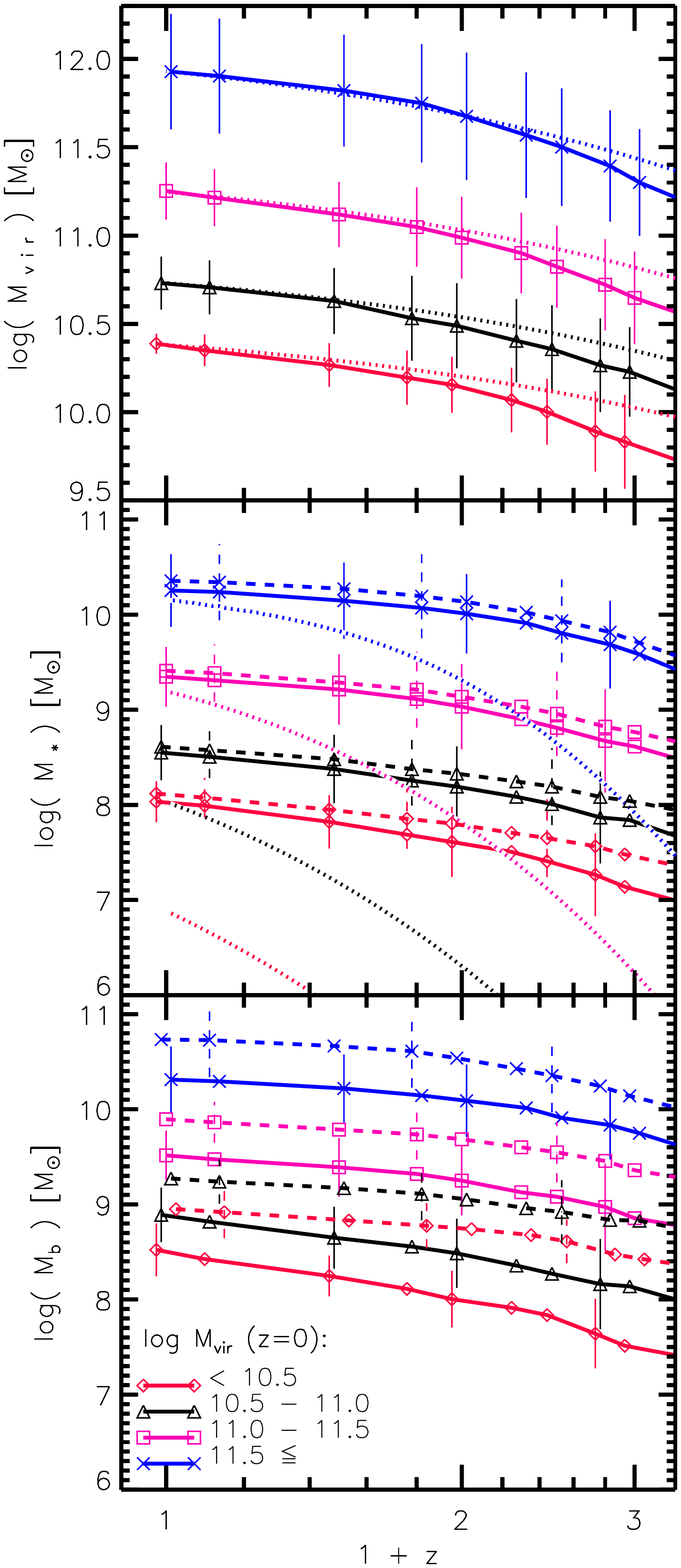}}
\vspace{-0.5cm}\hspace{0.5cm}\resizebox{7.8cm}{!}{\includegraphics{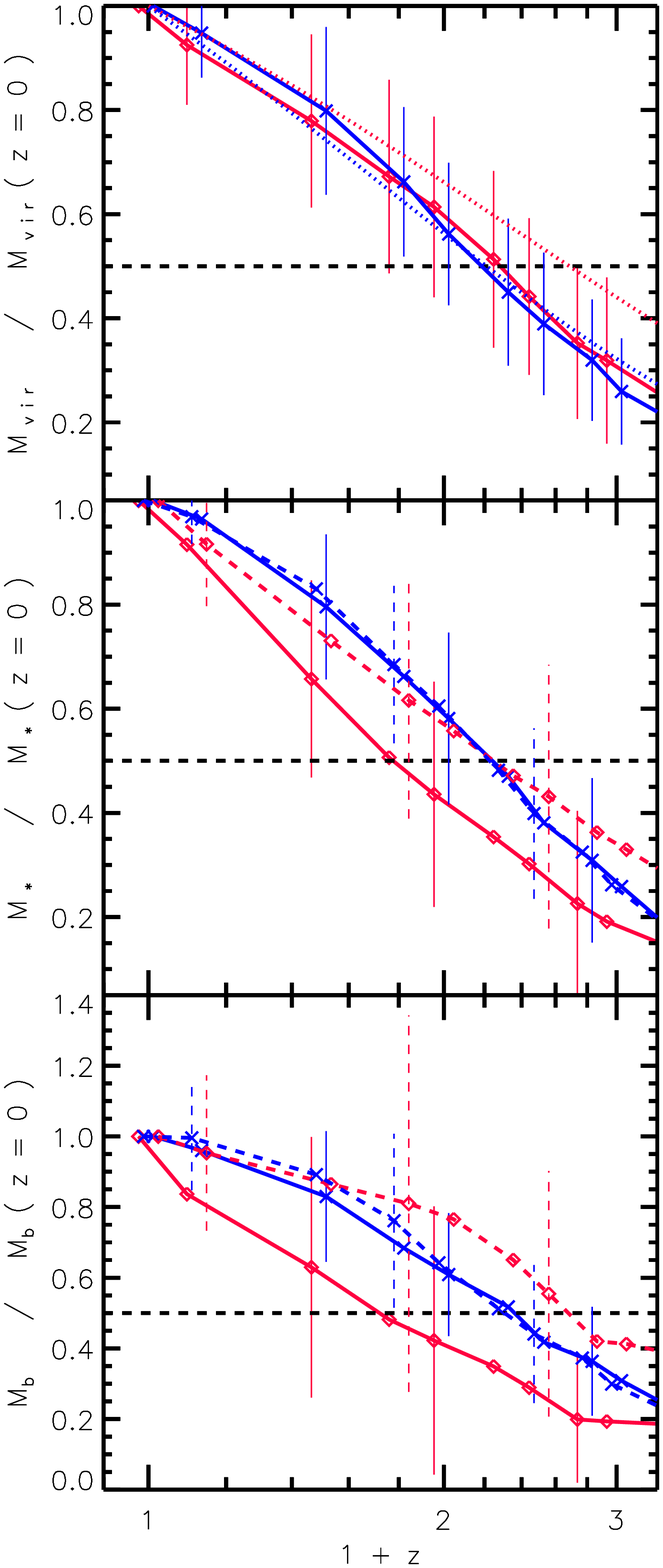}}
\end{center}
\caption[Mass aggregation histories]
{
Average halo (top panels), galaxy stellar (middle panels), and galaxy baryonic (bottom panels) 
mass aggregation histories for four $\log (\mh / M_{\odot})$ bins defined at $z = 0$ (solid lines):
$<10.5$ (red diamonds), $10.5-11$ (black triangles), $11-11.5$ (pink squares) and $\ge 11.5$ (blue crosses).
In the left panels, the average value of the progenitor mass at a given redshift
is calculated for each bin. In the right panels, the averages in each bin are calculated
considering the mass of each galaxy at a given redshift normalized to its final value at $z=0$.
For the sake of clarity, only the lowest and highest mass bins are shown in right panels (galaxies
in the other bins exhibit an intermediate behaviour).
Error bars represent $1\sigma$ population scatter.
In the case of the halo MAHs, the dotted lines show the results derived from eq. (2) of \citet{Fakhouri+2010}.
In the case of the stellar and baryonic components, together with the evolution of galaxy masses inside
\ropt, the corresponding evolution of masses inside the whole halo, up to $R_{\rm vir}$,
are also plotted (short-dashed curves). The dotted curves in the medium left panel corresponds 
to inferences from a toy parametric model constrained to fit observations (see Sect. \ref{comparison} 
and Appendix) and are associated to the same four mass bins used for the simulations, with  
$\log (\mh / M_{\odot})$ increasing from the lower to the upper dotted curve.
}
\label{fig:massagg}
\end{figure*}

In the left panels of Fig. \ref{fig:massagg}, we can appreciate the average virial
halo, galaxy stellar, and galaxy baryonic MAHs
for four different subsamples of galaxies defined at $z=0$
according to their log(\mh/\msun):
$< 10.5, 10.5-11.0, 11.0-11.5,$ and $\ge 11.5$. The 
error bars correspond to the standard deviations associated to each subsample 
at a given $z$. To avoid overplotting, the means and error bars are slightly shifted along the
horizontal axis and the error bars are not shown at all epochs.
The right panels portray the same information but the MAHs
were normalized to the corresponding present-day masses
in order to get more insight into the shape of the evolutionary tracks
and its dependence on mass.
For the sake of clarity, only the lowest and highest mass bins are shown in right panels (galaxies
in the other bins exhibit an intermediate behaviour).

In the upper panels, for comparison, we also show 
the mean halo MAHs from the Millennium 
simulations as fitted by \citet[][dotted lines]{Fakhouri+2010} and 
corresponding to the present-day masses shown for our 
simulations.  We obtain a general good agreement taking into account
that our box is much smaller than those of the Millennium simulations
and that those simulations include only collisionless (dark) matter. Systematic
differences can be appreciated at $z>1$,  implying a slightly later mass assembly 
in s230 than in the case of the Millennium simulations. These differences 
are more pronounced for smaller masses, which suffer more 
dramatic SN-driven outflows, specially at earlier epochs. 
Although with a large scatter, for both the Millennium simulations and s230, more 
massive systems tend to assemble their virial masses typically later than the 
less massive ones (hierarchical mass assembly or upsizing). However, in the case of s230, 
this trend is weaker, mainly due to the later assembly of simulated
low-mass systems. For instance, for the smallest mass bin at
$\log(\mh/\msun) <10.5$, the redshift at which \mh\ attains
50\% of its present-day value is 
$z_{\rm h,1/2} \approx 1.35$ (ranging from $z\approx 0.75$ to 1.81 for 
the 1$\sigma$ population), while at $\log(\mh/\msun) \ge 11.5$, $z_{\rm h,1/2}\approx 1.16$  
(ranging from $z\approx 0.80$ to 1.60 for the 1$\sigma$ population).

The solid lines in the middle panels show the average galaxy stellar MAHs. 
The slight upsizing trend with mass for the halo 
MAHs is reversed to a downsizing trend in the case of the stellar MAHs: 
galaxies in less massive halos assemble their present-day \ms\ 
slightly later than those in more massive ones, though the scatter is large. 
This trend can be better appreciated in 
the case of the normalized stellar MAHs (\ms($z$)/\ms($z=0$)). 
For the same two extreme ranges of virial masses mentioned above, 
$\log(\mh/\msun) <10.5$ and $\log(\mh/\msun) \ge 11.5$, the redshifts at which 
\ms\ attains 50\% of its present-day value are 
$z_{\rm s,1/2}\approx 0.80$
(ranging from $z\approx 0.4$ to 1.4 for the 1$\sigma$ population) and $z_{\rm s,1/2}\approx 1.30$
(ranging from $z\approx 0.8$ to 1.7 for the 1$\sigma$ population), respectively.
Hence,  simulations seem to be successful in predicting the observed 
downsizing trend
for \ms, albeit the tendency is weak.

\begin{figure}\hspace*{0.5cm}\\
\vspace{7cm}
\includegraphics{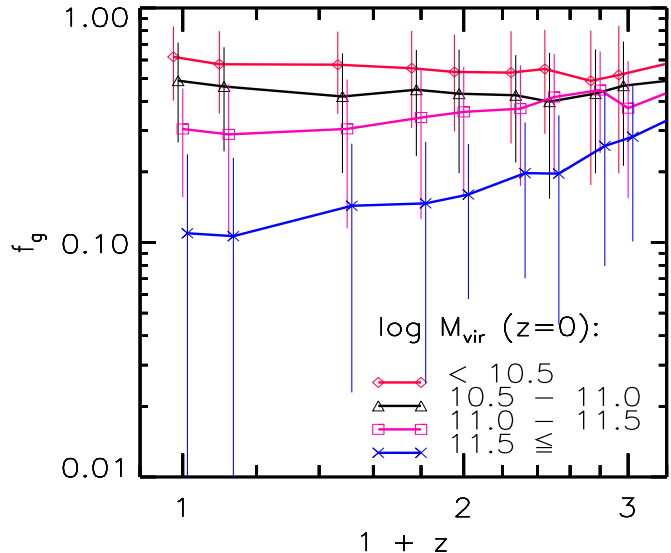}
\caption{
Evolution of the average gas fraction of simulated galaxies grouped into the same four 
halo mass bins defined at $z=0$ and shown in Fig. \ref{fig:massagg}. Error bars represent the  $1\sigma$ population scatter.}
\label{fig:fg-Ms}
\end{figure}

The galaxy baryonic MAHs do not differ significantly from stellar ones
(lower panels of Fig. \ref{fig:massagg}). The downsizing
trend obtained for \ms\ can be also appreciated in the case of \mb. However, differences in 
the absolute values are evident: while for the least massive systems, \mb\ is significantly larger 
than \ms\ at all available $z$ (roughly a factor of $\sim 3$),  
for the most massive ones, \mb\ is only slightly larger than \ms, being this difference 
somewhat larger at higher redshifts.
The mentioned trends are explicitly seen in Fig. \ref{fig:fg-Ms}, where the mean
evolution of \fgg\ is plotted for our simulated galaxies grouped 
according to their present-day halo masses.  As expected, less massive galaxies exhibit high percentages
of gas at all epochs while in the case of larger systems, \fgg\ significantly decreases
towards $z \sim 0$.

\subsection{Assembly of stars and gas inside virialized halos}

We can appreciate 
the MAHs for the whole stellar and baryonic components
inside \rh\ in the middle and lower panels of Fig. \ref{fig:massagg}, respectively (dashed lines).  
In the case of $M_*^{\rm vir}$, the mass fraction 
outside the central galaxy can be found in satellite galaxies and also in an extended stellar halo.
For massive halos, this fraction is small and it slightly increases with
$z$ from $\approx 20\%$ at $z\sim 0$ to $\approx 35\%$ at $z>2$. In the case of smaller halos, 
the stellar component outside central galaxies was significant at high $z$, decreasing towards 
lower $z$. 
 
Regarding the whole baryonic component, $M_b^{\rm vir}$ seems to be 
significantly larger than the baryonic mass contained inside central galaxies
at all epochs. These differences are higher in the case of less massive systems, 
in particular at high $z$. These results imply the presence of significant 
fractions of gas in simulated halos, with the greater percentages obtained for smaller systems
and at higher $z$ 
\citep[see also][]{deRossi+2010}. The MAHs represented by the evolutionary tracks
of $M_b^{\rm vir} / M_b^{\rm vir} (z=0)$ (lower right panel) are quite diverse 
(large error bars) specially for small halos, where there is a significant 
interplay between gas (re)accretion and feedback-driven outflows. 
It is clear that $M_b^{\rm vir}$ in some small halos decreases towards 
$z \sim 0$, probably due to strong outflows \citep[][]{deRossi+2010}. On average,  
the MAHs associated to $M_b^{\rm vir}$ seem to follow qualitatively the upsizing 
trend of the halo MAHs. However, less massive systems
tend to assemble $M_b^{\rm vir}$ earlier than the corresponding \mh.  
These trends are probably caused by a very efficient SN feedback 
that avoids late gas capture by the galaxy and promotes gas ejection from the halos. 
Nevertheless, our findings suggest that a significant amount of gas resides inside these halos, probably
supported by thermal pressure (hot-gas phase).
Part of this gas may cool later leading to an increase of the SFR in the galaxy.

\subsection{Evolution of the stellar and baryonic mass fractions}
\label{mass-fractions}

\begin{figure*}
\begin{center}
\resizebox{7.7cm}{!}{\includegraphics{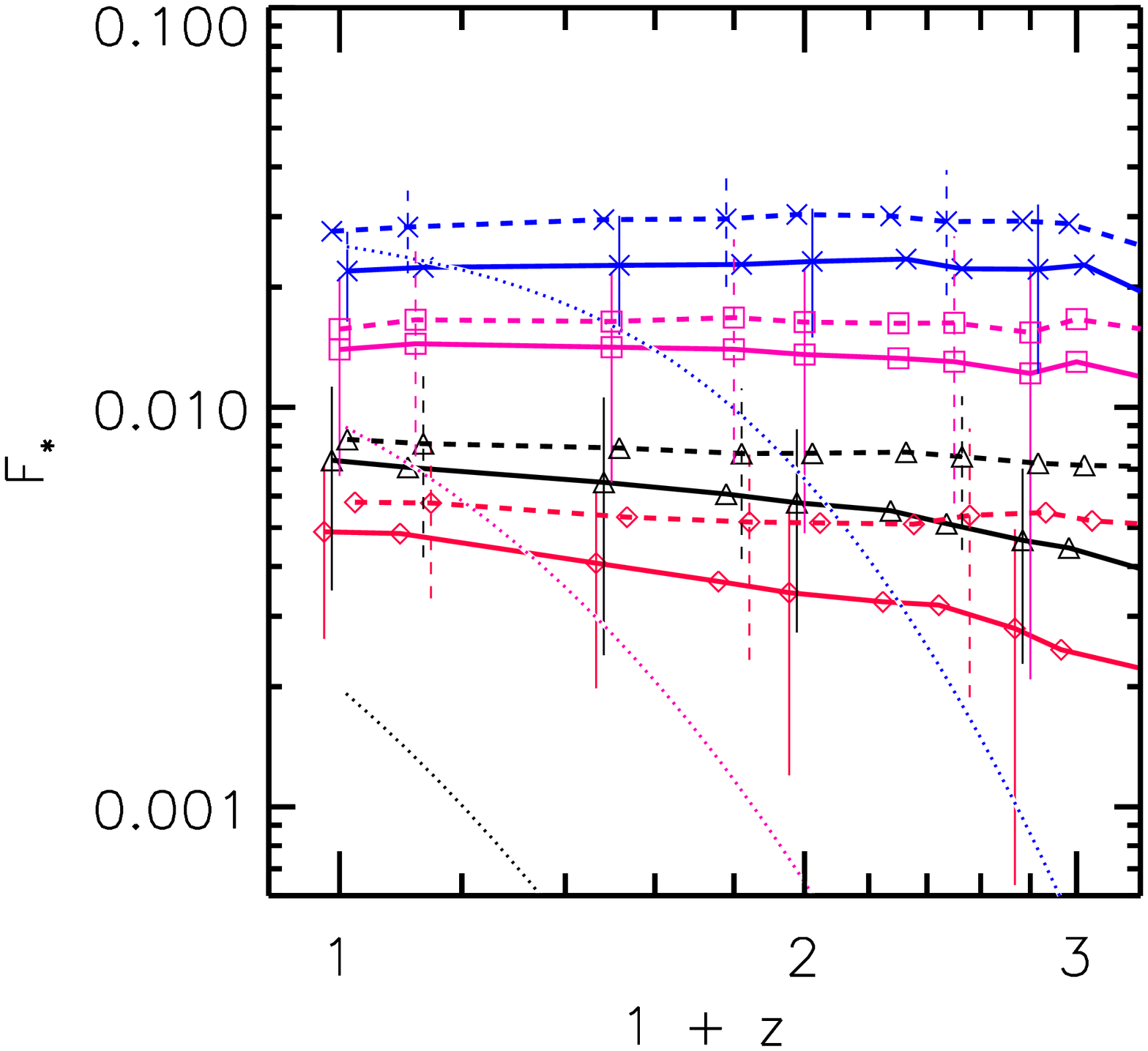}}
\vspace{-0.6cm} \hspace{0.5cm}\resizebox{7.7cm}{!}{\includegraphics{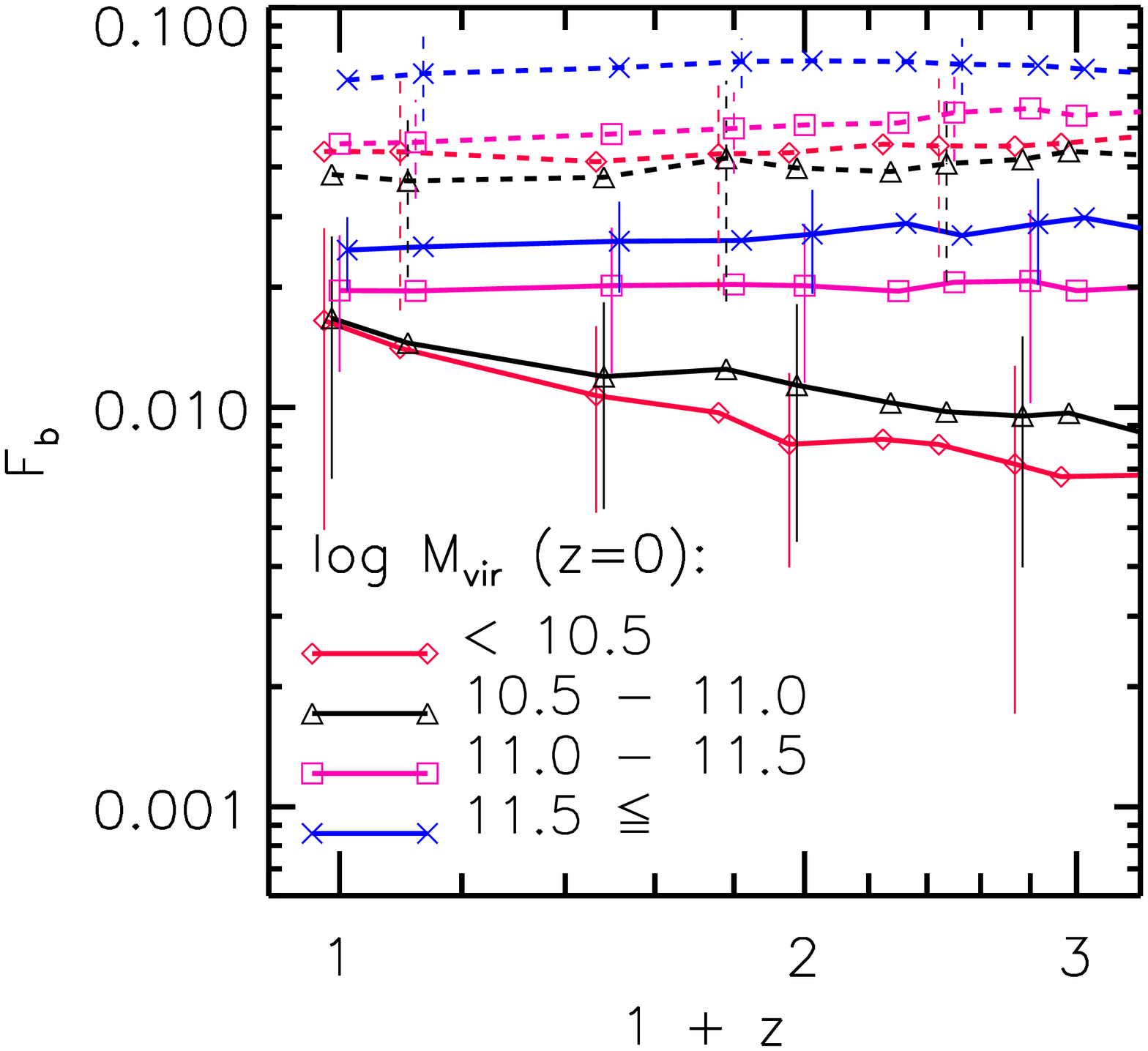}}
\end{center}
\caption[Mass aggregation histories]
{
Evolution of the average galaxy stellar (left panel) and 
baryonic (right panel) mass fractions (solid lines)
grouped into the same four 
halo mass bins defined at $z=0$ and shown in Fig. \ref{fig:massagg}. Error bars represent
 the $1\sigma$ population scatter. 
The stellar and baryonic fractions obtained by considering the whole mass enclosed inside \rh\
(\fst\ and \fbt, respectively) are represented with short-dashed lines. 
In the left panel,
the dotted curves depict the inferences derived from the toy
parametric model constrained to fit observations (see Sect. \ref{comparison} and Appendix)
and are associated to the same four mass bins used for the simulations, with
$\log (\mh / M_{\odot})$ increasing from the lower to the upper dotted curve.
}
\label{fig:massaggratio_v2}
\end{figure*}

The evolution of the average \fs\ and \fb\
(solid lines) can be appreciated in Fig. \ref{fig:massaggratio_v2} in the left and right panels, respectively. As in 
previous figures, the mean relations for four mass bins are shown, with the
error bars denoting the associated standard deviations. 
In both panels, the short-dashed lines show the results obtained for the stellar (\fst) 
and baryonic (\fbt) mass fractions measured inside \rh\ (instead of \ropt).

As can be seen, at all epochs, \fs\ is systematically lower in galaxies formed in 
present-day low-mass halos than in galaxies formed in more massive 
halos. In fact, the evolution
of \fs, though with a large scatter, is such that for low-mass galaxies,
\fs\ decreases towards higher $z$, while for more massive galaxies, \fs\ does not change significantly
up to $z\sim 2$ (for the most massive galaxies, \fs\ can be a bit larger at higher $z$). 
These findings suggest that for massive galaxies ($\ms>10^{10}\msun$), \ms\ tends to assemble 
at a similar rate than \mh, while for less-massive galaxies, \ms\ assembles later 
than \mh.  Regarding \fst\ (dashed lines), 
we did not obtain systematic variations with $z$ for any of the considered mass bins, 
suggesting that the total stellar and dark matter components
tend to increase at a similar rate inside \rh.

In the case of galaxies in massive halos, the evolutionary tracks of \fb\ 
(right panel of Fig. \ref{fig:massaggratio_v2}, pink and blue curves)
follow closely those obtained for \fs. 
For less massive halos, as noted before (see Figs. \ref{fig:fgasvsz} and \ref{fig:fg-Ms}), 
the gas-phase tends to dominate the baryonic 
component of simulated galaxies, being $\fb > \fs$. 
In particular, for galaxies in the lowest-mass bins, \fb\ is not only much higher than \fs\ 
but it increases slightly faster than \fs\ since $z\sim 1$, probably evidencing
the re-infall of gas ejected from the systems at early times.

Regarding \fbt, 
we do not detect systematically variations with $z$, on average. Therefore, the baryonic content 
inside \rh\ follows 
roughly the MAH of the halo (specially in the case of the most massive bin), 
as discussed before. 
In fact, since a similar trend was also obtained for \fst, the different components
(gas, stars and dark matter) inside \rh\ seem to increase at the same rate in these simulations,
at least since $z \sim 2$.
On the other hand, the percentage of baryons (mainly gas) in the halos is of the order or larger 
than the baryonic fraction contained in the central galaxies. 
At $z=0$, for instance, the average \fbt\ is $\sim 2-2.5$
times larger than \fb\ in the whole simulated mass range. Hence, 
outside the central galaxies there is,
on average, an amount of mass similar or up to $\sim 1.5$ times the baryonic mass of these
galaxies. In the case of the stellar component, 
\fst\ is only up to $\sim 1.2$ times larger than \fs.
Thus, most of the baryonic matter outside central galaxies is actually in the gas-phase.
At higher redshifts, the differences between \fb\ and \fbt\ increase for
less-massive galaxies. At $z=3$, for example, galaxies residing in the lowest-mass present-day
halos had $\sim 7$ times more baryons outside the galaxy than within it, being most of these
baryons part of the hot-gas component. 

\begin{figure*}
\begin{center}
\resizebox{8.5cm}{!}{\includegraphics{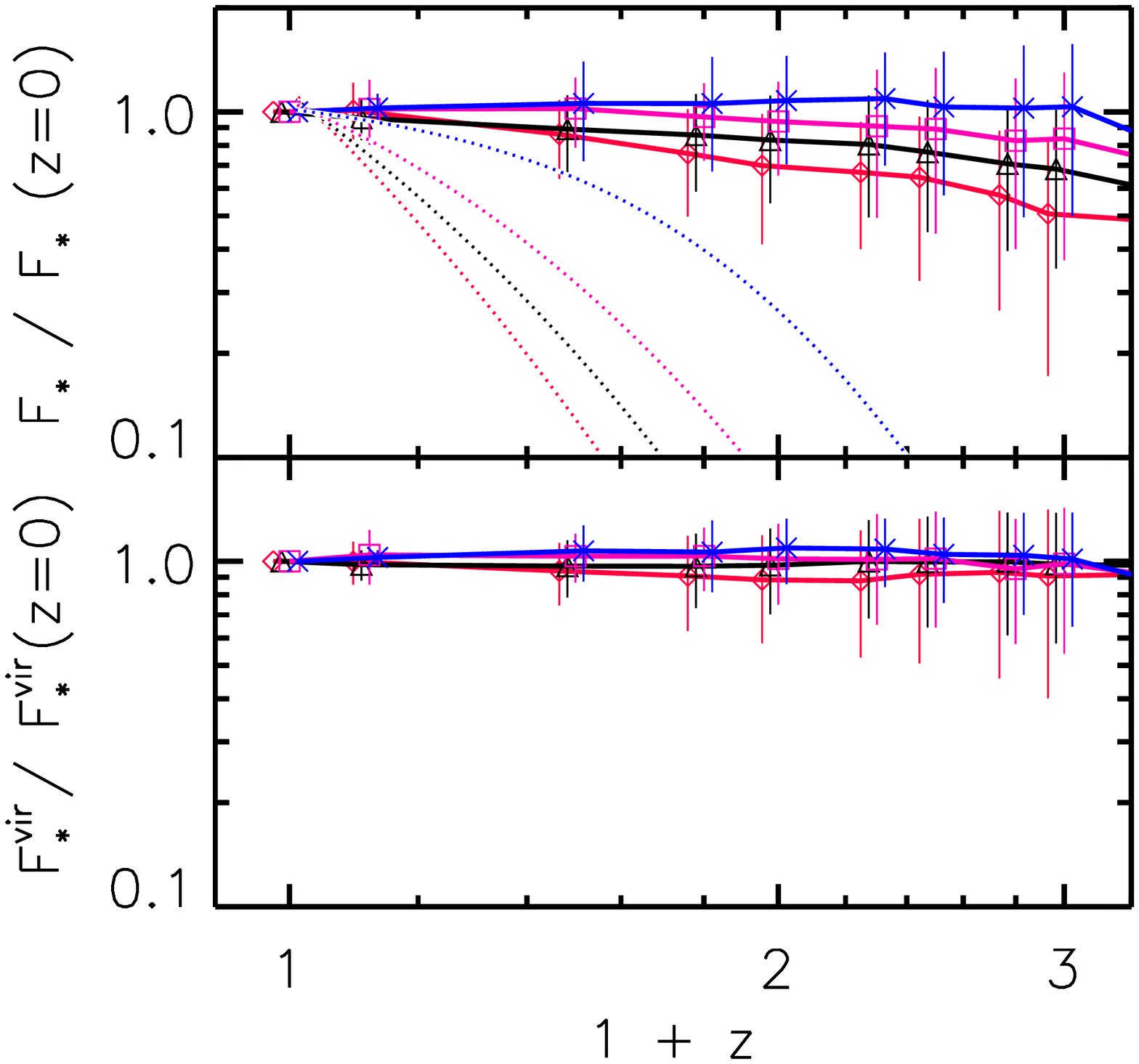}}
\vspace{-0.6cm}\resizebox{8.5cm}{!}{\includegraphics{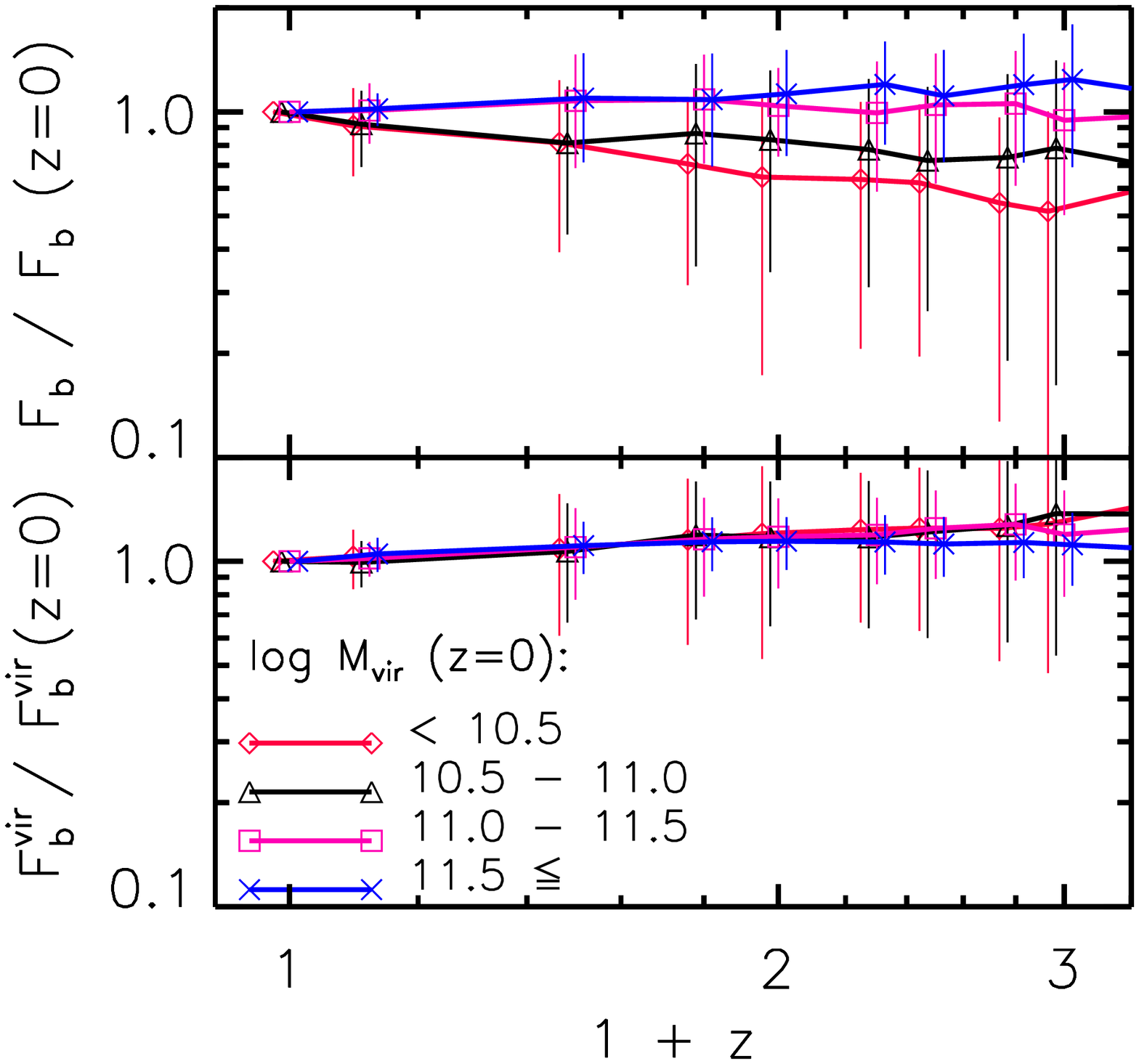}}
\end{center}
\caption[Mass aggregation histories]
{
Evolution of the same stellar (left panels) and baryonic (right panels) mass fractions shown in 
Fig. \ref{fig:massaggratio_v2} but normalized to their present-day values in such
a way that the patterns of the mass-dependent evolutionary tracks can be more easily compared.
Error bars show the corresponding standard deviations.
The upper panels show results for galaxy masses enclosed by
\ropt, while lower panels correspond to total masses inside \rh. 
The averages were derived considering the mass fraction of each galaxy at 
a given $z$ normalized to its value at $z=0$.
In the upper left panel, the dotted curves indicate the inferences from the toy
parametric model constrained to fit observations (see Sect. \ref{comparison} and Appendix)
and are associated to the same four mass bins used for the simulations, with
$\log (\mh / M_{\odot})$ increasing from the lower to the upper dotted curve.

}
\label{fig:Fratio}
\end{figure*}

Figure \ref{fig:Fratio} summarises the evolution of 
\fs, \fst, \fb\ and \fbt\ as a function of the present-day halo mass. 
In this case, the individual mass fractions 
were normalized to their present-day values 
before averaging over the different mass bins
so that, 
the patterns associated to the different evolutionary histories can be more easily compared.

Results for \fs\ are presented in the upper left panel.  Although the scatter is large, 
the stellar MAH of galaxies
is clearly dependent on \mh: as the mass of galaxies decreases, 
the stellar component tend to assembles later than the halo.
For the smallest systems,
\fs\ increases, on average, by a factor of $\sim 2$ since $z \sim 2$ to $z \sim 0$, while for the 
largest ones, \fs\ does not change significantly or slightly decreases with time in some cases.
Hence, for more massive systems, the  
stellar MAHs seem to follow closely the MAHs of the halos.
In the lower left panel, we can  appreciate the evolution of \fst.  As can be seen, 
for the whole component inside \rh,
the patterns associated to the mean evolution of the stellar-to-halo mass ratio 
are approximately scale-independent, 
exhibiting also no significant evolution since $z\sim 2$.

According to the right panels of Fig. \ref{fig:Fratio}, the baryonic-to-halo MAHs
of galaxies seem to be dependent on scale but with a much larger
scatter than the one obtained for the stellar-to-halo MAHs. For the smallest systems, the average \fb\ 
increases by less than a factor of $\sim 2$ since $z \sim 2$ while for the largest ones,
\fb\ does not change significantly or slightly decreases with time. 
In the case of \fbt, though with a large scatter, its evolutionary history seems to show an opposite trend
to the one obtained for \fs\ or \fb: for galaxies with lower \mh, 
\fbt\ tends to exhibit a more significant decrease with time.
These findings suggest that in these simulations, many 
low-mass halos tend to lose baryons with time, but we have seen that
there is also a non-negligible percentage of low-mass 
systems that do not experience significant outflows and/or re-accrete baryons lately.

\subsection{Comparison with empirical inferences}
\label{comparison}

With the aim of comparing the average MAHs derived from simulations and 
empirical inferences,
we use  a simple toy parametric model of the average halo and stellar mass growth constrained by
observations \citep[][see the Appendix for a more detailed description]{Gonzalez-Samaniego+2012}.
In this toy model, the average evolution of \ms\
inside the growing halos describes, by construction, the empirical 
sSFR--\ms\ and \fs--\mh\ mean relations at different redshifts, which can be considered isochrones
of the evolutionary tracks.
As discussed in the Appendix, 
the increase of \ms\ is driven mainly by in-situ SF but, in order to obtain agreement with the 
empirical relations, a small contribution of ex-situ stellar mass acquisition (dry mergers) is necessary for high 
masses.  
It is worth noting that this toy model does not attempt to include physical prescriptions
for the evolution of baryons; it is just constrained to reproduce observations. Similar approaches were
presented previously in the literature \citep[FA10,][]{Conroy+2009a, Leitner2012} and, in all cases, 
the most general trends are 
consistent to the ones derived from our toy model. Furthermore,  these trends are in good
agreement, at the masses and redshifts when 
a comparison is possible, with archaeological inferences (see the Appendix).
In Sec. \ref{sec:results}, the results of this toy model were 
used as descriptions of observations and
compared with 
the predictions of our numerical simulation. In this section, we will
extend this comparison to the average
MAHs of galaxies.

As shown before, in the left middle panel of Fig. \ref{fig:massagg}, we can appreciate the  
toy model tracks associated to \ms\ (dotted lines) for galaxies with present-day 
log(\mh/\msun) = 10.30, 10.75, 11.25, 
and 11.75 (from bottom to top), which are representative of the mass bins used in the simulations. 
As the mass decreases, the stellar MAHs of simulated galaxies exhibit more
significant deviations from the toy model tracks. As 
anticipated, in the simulations, \ms\ tends to assemble earlier than what
the toy model indicates, being the difference systematically 
larger for lower masses. 
Regarding the halo mass assembly, the toy model is based on the mass aggregation rates 
given in \citet{Fakhouri+2010}, which are shown with dotted curves in the upper panels of Fig. \ref{fig:massagg}.

It should be noted that 
the toy model is constrained by observations only above $\sim 10^{9}$ \msun\ and
the lowest-mass tracks are actually extrapolations. Studies of the past SF history of relatively 
isolated nearby dwarf galaxies by means of
color-magnitude diagrams obtained with the Hubble Space Telescope (HST) show that low-mass systems (specially below  $\sim 10^{8}$ \msun) typically assembled 
most of their stars at $z>1-2$ \citep[][]{Weisz+2011b}. 
The present-day sSFR of these field dwarfs seems to be lower than the extrapolation to 
lower masses of the trends in \citet[][Fig. \ref{fig:sSFRvsM}]{Salim+2007}.
On the other hand, studies based on UV analysis, assign
systematically higher SFRs to dwarf galaxies, in better agreement with \citet[][e.g., \citealp{Lee+2011,Huang+2012}]{Salim+2007}. 
Above $\sim 10^8$ \msun, as the mass decreases,
the stellar mass assembly is systematically delayed.
On the contrary, below $\sim 10^8$ \msun, this trend seems to be
interrupted, with the stellar component of small galaxies commonly assembling very early \citep{Leitner2012}. 
Therefore, the evolutionary tracks inferred with our toy model should be taken with caution at
\ms\ below $10^8$ \msun\ (lowest dotted line in Figs. \ref{fig:massagg}, \ref{fig:massaggratio_v2}, and \ref{fig:Fratio}).

In the left panels of Figs. \ref{fig:massaggratio_v2} and \ref{fig:Fratio}, 
we can appreciate that the evolution of \fs\ inferred with the toy model 
deviates from the one predicted by the simulations. 
As the mass decreases, the toy model suggests a more significant 
delay in the stellar mass assembly with respect to the halo one.
For galaxies in halos with present-day masses of $\sim 10^{12}$ \msun, 
the stellar mass fractions associated to the toy model decrease
$\sim 3$ times more towards $z \sim 1$ than what simulations suggest, 
while this difference increases roughly up to 20 times for galaxies in halos of $3-10 \, \times 10^{10}$ \msun\
(compare the corresponding dotted and solid curves in Fig. \ref{fig:Fratio}). 
Part of this discrepancy is caused by the different halo MAHs associated to the toy model
and the simulations: the simulated \mh\ decays faster with $z$ than the dark matter MAHs given in
\citet[][see Fig. \ref{fig:massagg}]{Fakhouri+2010}.

The delay in the stellar mass assembly of smaller galaxies could 
also depend on the observational data used to constrain the toy model. In particular, \citet{Yang+2012}
have recently reported new inferences regarding the evolution of the \ms--\mh\ relation: for one of the cases 
analysed by those authors at low masses, at a given \mh, \ms\ decreases faster than what the \ms(\mh,$z$) function given 
by FA10 implies.  
In our toy model, the tracks that would fit the \ms(\mh,$z$) function reported by \citet{Yang+2012} 
lead to steep sSFR--\ms\ relations at low masses. Thus, the delay in the \ms\ growth
of smaller galaxies could be even more dramatic than what we have discussed here.
On the other hand, \citet{Behroozi+2012} obtained new constraints on the \ms(\mh,$z$) function.
According to the latter work, the \ms(\mh,$z$) function does not change significantly with $z$ in such a way 
that for smaller galaxies, the stellar mass growth 
is shallower than what we have inferred here, %
leading to a better agreement with our simulations but yet with differences.

Finally, although the trend is weak, it is encouraging that the SN feedback model 
used in this work
is able to reverse the upsizing trend of the halo MAHs to a downsizing trend
in the case of stellar MAHs.  
In the next section, we discuss about possible additional physical prescriptions
that could be implemented in the simulations to tackle the issue of the early
stellar mass assembly and obtain a 
better agreement with empirical inferences.

\section{Discussion}
\label{phys-processes}

As we mentioned in Sec. \ref{sec:intro}, this work does
not attempt to analyse in detail the particular predictions of the
subgrid model implemented in these simulations.
We aim mainly to discuss about the general trends which are shared
with other numerical works based on current models and simulations
of structure formation.
As discussed before,
recent results obtained by re-simulating selected objects at high resolution (see Sec. \ref{sec:intro})
lead to reasonable consistency with observations.
Although the resolution of our simulations is lower than in those studies, both approaches
predict similar trends in general, with the advantage that our simulated sample include
a larger number of systems.  Hence,  we are able to analyse statistically the evolutionary
histories of galaxies covering a more significant dynamical range. 
Exhaustive explorations of different subgrid schemes and physical parameters were carried out recently 
in the literature, showing how the properties and evolution of simulated galaxies may change 
according to the subgrid assumptions \citep[e.g.,][]{Saitoh+2008, Colin+2010, Dave+2011,
Faucher-Giguere+2011,Hummels+2012, McCarthy+2012,Scannapieco+2012}. 
The simulation analysed here is actually part of such studies, having
been its subgrid physics and parameters chosen to produce nearly realistic 
galaxies in what regards global structural and dynamical properties \citep[][]{deRossi+2010, deRossi+2012}.

\subsection{Subgrid physics in the simulations}

The modelisation of SF-driven outflows is crucial for reproducing the flattening of the GSMF at the 
low-mass end or, similarly, for predicting the decrease of \fs\ for lower \mh. 
For example, \citet{Oppenheimer+2010} showed how the GSMF changes according to different 
feedback models implemented in their TreeSPH GADGET-2 simulations. In all these models, the feedback 
is directly related to the SFR and is included as a kinetic energy added to gas particles.
These particles are temporarily hydrodynamically decoupled in order to provide a low resistance avenue for SN feedback to escape
out of the galaxy. \citet{Oppenheimer+2010} found that 
the gas ejected and lately re-accreted by the galactic systems is a key mode of galaxy growth. If the gas
acquired by the galaxies through this mode is artificially suppressed, then the (unrealistic) 
GSMFs obtained at $z=0$ are similar in spite that the feedback models are different. But, when taking into 
account the re-accreted gas, then the GSMF depends sensitively on it. 

\citet{Oppenheimer+2010} reported that a better agreement with the observed GSMF is obtained 
when rather than an "energy-driven" wind, a "momentum-driven" wind is introduced \citep{Murray+2005}. 
However, the re-accretion mode is mass-dependent, increasing its efficiency
as the mass increases (for massive galaxies, a local fountain effect is obtained). The 
parameters of the momentum-driven winds can be tuned to reproduce roughly the low- and
intermediate-mass regions of the GSMF due to the late ($z\lesssim 2$) differential gas re-accretion. 
However, at the high-mass end, too massive galaxies are obtained because of the fast 
re-accretion exhibited during the evolutionary history of these galaxies. 

These results are similar to those obtained by \citet{Firmani+2010a} by means of
disc galaxy evolutionary models in the context of the \lcdm\ scenario. 
They showed that the \fs--\mh\ relation (associated to the GSMF) at low and intermediate 
masses can be reproduced by the models when efficient momentum-driven outflows and 
differential re-accretion depending on the infall rate (environment) are included.  However, 
the models in this case show a sSFR that increases with \ms, being close to present-day observations
at intermediate masses ($\ms\sim 3-6\times 10^{10}$ \msun),
but systematically lower than observations for lower \ms.  This behaviour can be explained by the 
early mass loss and the decreasing efficiency of gas re-accretion for lower masses. The 
latter result has been also found by \citet{Dave+2011}, who used TreeSPH numerical simulations similar to those of 
\citet{Oppenheimer+2010}.  These authors show that an outflow model following 
scales expected for momentum-driven winds broadly matches the observed galaxy evolution 
around \mstar\ since $z = 3$ to 0, but it fails at higher and lower masses.

The SN-driven feedback model implemented in our simulation is physically more self-consistent 
than those based on kinetic energy input and temporary hydrodynamical decoupling in the sense 
that the model is tied to a multi-phase treatment of the gas components in the ISM. Note that the 
energy injected by SNe to the gas particles is thermal and it depends on the thermodynamic 
properties of the particles. This energy can promote particles from the cold/dense phase 
to the hot/diffuse phase, influencing thus the SFR. 

The SN feedback modelled in this way has therefore a relevant influence
both locally, affecting the gas properties and the SFR, and globally, producing pressure-driven
large-scale gas and metals outflows. The latter effect is the one that produces the flattening 
of the GSMF at the low-mass end or the decrease of \fs\ for lower \mh\ (see above). 
Note that this effect can also produce a systematic 
decrease of the sSFR at late epochs for lower masses because smaller galaxies lose 
more efficiently their gas and are less susceptible to re-accrete it lately \citep{Firmani+2010a,Dave+2011}. 
In this context, it is relevant to remark that in our simulation, the present-day sSFR 
does not decrease towards lower \ms\ (Fig. \ref{fig:sSFRvsM}), on the contrary, it increases with a slope close to 
observations. This success is due to the local effects of our thermal feedback model: a 
significant fraction of the left-over gas in small galaxies is maintained in a hot phase in the disc 
and halo, being able to cool lately and produce a sustained SF activity. According to 
Fig. \ref{fig:fgasvsz}, as the mass of simulated galaxies decreases, higher gas fractions
are obtained. In addition, the large amounts of gas in the halos of small galaxies tend to
decrease with time contributing to the increase of the galaxy stellar and gas mass fractions 
(see Fig. \ref{fig:massaggratio_v2}).
For the largest galaxies, the gas in the halo does not exhibit significant variations with time. 

All these findings are consistent with the presence of two different
thermodynamic regimes in these simulations as described in detail by \citet{deRossi+2010}.
In the case of smaller galaxies, the virial temperatures are lower
and, therefore, SN heating is more efficient at promoting gas from the cold to the hot phase.
However, the cooling times of these systems are shorter than the dynamical
times and the hot gas can return to the cold phase on short time-scales.
Therefore, for low-mass galaxies, SN feedback leads to a self-regulated cycle of
heating and cooling strongly influencing the SF activity of these
systems. In the case of massive galaxies, the hot phase is established at a
higher temperature and, hence, SN heating cannot generate an efficient transition of
the gas from the cold to the hot phase; meanwhile, the cold gas remains
available for SF. In addition, the cooling times for larger
galaxies get longer compared to the dynamical times and the hot gas is able to
remain in the hot phase during longer time-scales. Hence, SN feedback is not
efficient at regulating the SF in massive galaxies.
As shown by \citet{deRossi+2010}, in this model, this transition from an efficient to an inefficient
cooling regime for the hot-gas phase produces
a bend of the stellar Tully-Fisher Relation in good agreement with observations.

In spite of this partial success, in Sec. \ref{sec:results} and \ref{comparison}, we have
seen that the sSFR--\ms\ and \fs--\mh\ correlations, and the MAHs of the low-mass 
simulated galaxies still exhibit discrepancies with some observational inferences. 
Although in our  simulation the upsizing trend of dark 
matter seems to be reverted to a downsizing trend in the case of stellar mass, this behaviour is still weaker than what
observations suggest.  In the next section, we discuss about different possibilities
to tackle these issues.

\subsection{What should be improved?}

Would the increase of the feedback model efficiency help to solve the above-mentioned issues? 
The parameters of our feedback model were constrained to reproduce
properties of MW sized galaxies \citep{Scannapieco+2008}, being the typical outflow
velocities generated in simulated galaxies consistent with observational inferences
\citep{Scannapieco+2006,Scannapieco+2008}. The fraction of SN energy and metals injected into the
cold phase, $\epsilon_c=0.5$, could hardly be increased. As discussed
in \citet{Sawala+2011}, for dwarf galaxies, this efficiency should be decreased in order to be 
consistent with the observed mass--metallicity relation. Moreover,
the feedback-driven outflows seem to be very efficient in our simulation: halos that today
are smaller than $\mh\sim 10^{11}$ \msun\ have inside them only around 25\% of the universal
baryon fraction since $z\sim2$, while the largest halos in our simulation ($\mh> 3\times 10^{11}$ \msun) 
have also small baryonic fractions ($\approx 35-40$\% of the universal one, see Fig \ref{fig:massaggratio_v2}).  

In fact, the comparison with observational inferences shows that besides 
the increase of the ejected mass for smaller galaxies, a delay in the SF process is also necessary. 
A more complete treatment including multiple feedback processes on different
scales could work in this direction. For instance, in addition to the SN wind shock heating,
\citet{Hopkins+2012} considered also the momentum deposition from radiation pressure, 
SNe, stellar winds, the photoheating of HII regions, and other processes.
These authors show that the effects of all these feedback processes affect
in different ways the galactic winds as well as the properties of the simulated 
galaxies, depending on their masses and SF regimes. 
On the other hand, \citet{Puchwein+2012}
claim that by tuning appropriately the energy-driven kinetic feedback in their 
SPH simulations, the observed \fs--\ms\ and sSFR--\ms\ relations, and other 
properties could be reproduced. The
key ingredients in their feedback model are the assumptions that the wind velocity 
depends on the galaxy potential well and that the loading factor is 
proportional to this velocity. 

During the refereeing of this paper, a preprint by \citet{Aumer+2013} appeared, where
the authors discuss some improvements on the \citet{Scannapieco+2008} subgrid
physics used here. In the new proposed scheme, the effects of radiation pressure from massive young
stars on the ISM are included, helping to reproduce disc galaxies with small bulges, sizes, star formation
rates, among other properties, in global agreement with observations at $z=0-3$.  For higher $z$, higher
star formation efficiencies are still present, suggesting the need for new physical processes to be included.

Another avenue of exploration in simulations is related to the SF process. For 
example, \citet{Krumholz+2011} and \citet[][]{Kuhlen+2012} have proposed that 
the formation of molecular gas, $H_2$, in galaxies could be delayed in low-mass/low-surface 
brightness galaxies because they have lower metallicities at earlier epochs. Only after a 
threshold metallicity and gas surface density are fulfilled, neutral hydrogen transforms
efficiently into $H_2$ in the high-density, cold regions, being the smaller galaxies subject to
this process at lower
redshifts. 

A recent SPH cosmological simulation of a dwarf
galaxy that includes an explicit method for tracking the non-equilibrium abundance of $H_2$,
shows that the dwarf has a larger gas fraction and higher SFR at later times than the
simulation without the $H_2$ treatment \citep{Christensen+2012}.

Finally, as discussed before, our results seem not to be significantly affected by resolution issues. 
When comparing results derived from s230 with those obtained for the higher-resolution
simulation s320 (available only at $z \ge 2$), similar trends were found.
In Figs. \ref{fig:sSFRvsM} and \ref{fig:fivsMh}, the dotted lines at $z = 2$ indicate the trends
obtained by using s320. We can appreciate that the relations associated to s230 and s320
are very close. Galaxies in s320 seem to have been slightly more efficient in assembling their stellar
component than systems in s230 but the differences are not significant.  
Even at higher redshifts ($z > 2$), the trends exhibited by s230 
remain very similar to those obtained for s320, 
though the small aforementiononed differences seem to be more evident at these $z$
(see Fig. \ref{fig:sSFR-s320}).  Regarding the cosmic SFR histories of both runs (Fig. \ref{fig:cosmicSFR}), 
we see that, in general, s230 and s320 lead to similar trends exhibiting also no significant
differences. These findings are consistent with the results of \citet{deRossi+2010},
who found that the dynamical properties of galaxies in s230 seem to be robust against
numerical artefacts (the reader is referred to that paper for more details).

\section{Conclusions}
\label{sec:conclusions}

A whole population of low-mass galaxies was simulated
with the aim of studying their stellar, baryonic, and dark halo mass
assembly in the context of a \lcdm\ cosmology and 
current SF and feedback models. The box of a comoving 14.3 Mpc side-length
represents an average (field) region of the universe. Only galaxies resolved
with more than $N_{\rm sub} = 2000$ particles ($\mh\gtrsim 10^{10.3}$ \msun) 
were analysed. The (rare) most massive 
halos in the simulated volume have masses of $\mh\approx 1-3\times 10^{12}$ \msun.
The simulations were performed by using the SPH GADGET-3 code 
with a multiphase model for the ISM 
and a thermal SN feedback scheme. Since our simulation reproduces a whole 
population of galaxies, we were able to statistically study their properties at 
different epochs, as well as average evolutionary trends that reveal how 
the population was assembled. Most of the simulated systems in the box correspond to central
galaxies.
In order to explore very high redshifts, a higher-resolution run with the
same initial conditions but available only at $z \ge 2$ was used.
 
Our main conclusions can be summarised as follows:

$\bullet$ In the simulations, the sSFR tends to increase as \ms\ decreases (downsizing
in sSFR) but, at very high $z$, this relation becomes flat and even inverts its slope at 
certain early epochs. 
The simulated and observational trends generally agree, though
the simulated sSFRs tend to be lower than those observed,
specially at low redshifts.  In addition, most of the simulated galaxies exhibit sSFRs lower
than the corresponding
to a SFR which was constant in the past.  Hence, simulated galaxies seem to be already (since $z\sim 1$) in a passive 
regime of mass growth due to SF, specially at greater masses. 
We do not find significant differences between the sSFR--\ms\ relations of central and satellite systems.

$\bullet$ At all analysed redshifts, \fs\ significantly decreases towards lower \mh, while
\fb\ also exhibit a similar behaviour but in a shallower way,
specially towards $z\sim 0$. 
This behaviour is caused by the higher \fg\ obtained for 
small galaxies since early epochs. For larger systems, on the other hand, \fg\ decreases with time.   
At $\mh\sim 10^{12}$ \msun, $<\fs>\approx 0.022-0.025$ and $<\fb>\approx 0.025-0.028$
since very high redshifts, being much lower than the universal baryon mass fraction. 
At $z\sim 0$, halos with masses $\gtrsim 10^{11}$ \msun\ exhibit similar values of \fs\ 
than those derived from empirical inferences. In the case of less-massive halos, 
simulated galaxies show slightly larger values of \fs\ than the associated 
to these inferences.
Regarding the evolution of the \fs--\mh\ relation, significant differences are obtained
between simulations and empirical inferences at low masses.
According to these inferences, at a given \mh, \fs\ tends to decrease as $z$ increases, 
while for simulated galaxies, the \fs--\mh\ relation 
does not exhibit significant variations since $z\sim 2$, evidencing an earlier stellar mass 
assembly in the case of simulations.

$\bullet$ At $z>2$, the most massive galaxies 
($\ms\gtrsim 3-5\times 10^9$ \msun) 
in the high-resolution counterpart of our simulations 
were compared with available observational data for these masses.
Their sSFRs are close to observations, though on average, the latter tend to exhibit higher values even 
at $z \sim 5$. The number density of simulated galaxies with masses between log(\ms/\msun)$\approx 9.25$ 
and 9.75 is somewhat higher than some observational determinations, while the cosmic
SFR history roughly agrees with observations up to $z \sim 4$. 

$\bullet$ In the most massive present-day halos, the average galaxy stellar mass only slightly 
increases since $z\sim 2$, while in smaller halos, the late mass growth develops faster. Although
with a large scatter, the upsizing (hierarchical) trend of halo assembly seems to be (moderately) 
reverted to a downsizing trend in \ms. A similar behaviour applies to the baryonic 
MAHs of galaxies. According to these findings, smaller galaxies 
exhibit a more significant delay of their active baryonic and stellar mass growth 
with respect to the halo MAH.
On the other hand, regardless of the mass,
the total stellar and baryonic mass inside the virial radius show similar assembly 
histories to the ones obtained for the corresponding halos. 
In particular, at $z=0$, the baryonic mass inside the halos (mostly in the hot-gas phase) is a factor of $\sim 1.5$ greater 
than the mass contained in the associated galaxies.
At $z=2$, this factor increases to $\sim 7$ for the lowest-mass systems.

$\bullet$ The stellar mass assembly
inside growing \lcdm\ halos
inferred with a simple toy model constrained to reproduce the empirical sSFR($\ms,z$) and \ms($\mh,z$)
relations, shows a downsizing trend. These stellar mass tracks evolve faster
at late times than those obtained in our simulations, even for galaxies in present-day halos with
$\mh\sim 10^{12}$ \msun.  
Despite that in the simulations, smaller
systems tend to delay their stellar mass assembly with respect to the halo assembly, 
these trends are still weaker
than those implied by current observational studies of low-mass galaxies
($10^9\lesssim \ms/\msun\lesssim 3\times 10^{10}$). 

In conclusion, we have found that for our simulated galaxy population, less-massive systems
exhibit a more significant delay of their active stellar mass growth with respect to the halo mass
assembly (downsizing in sSFR). However, this trend is still weaker than what empirical inferences
suggest. 
The multiphase ISM and thermal SN-driven feedback model implemented in these simulations
help to produce the downsizing in sSFR but there is little room for a more efficient (local and global) 
SN-driven feedback to improve the agreement with observations. 
Other feedback processes (radiation pressure due 
to massive stars, stellar winds, HII photoionization, etc.) could work in this direction. The weak delay 
in the active SF process for smaller galaxies could be also suggesting the need for the inclusion 
of additional subgrid SF physics or other processes related to the formation of $H_2$.  
Finally, it is worth mentioning that, at $z\sim0$, the observational inferences regarding the 
stellar mass fractions and mass assembly 
histories for galaxies smaller than $\ms\sim 5\times 10^9$ \msun\ are yet controversial.
Theoretical works as the present one could serve as a guide for these observational studies.

\section*{Acknowledgements}

We thank Octavio Valenzuela for useful comments on an early version of this work. 
We acknowledge a CONACyT-CONICET (M\'exico-Argentina) bilateral grant for partial funding.
V.A. and A. G. acknowledge PAPIIT-UNAM grant IN114509 and CONACyT grant 167332.  
A.G. acknowledges a Ph.D. fellowship provided by CONACyT.  
M.E.D.R., P.T. and S.P. acknowledge support from the  PICT 32342 (2005),
PICT 245-Max Planck (2006) of ANCyT (Argentina), PIP 2009-112-200901-00305 of
CONICET (Argentina) and the L'oreal-Unesco-Conicet 2010 Prize.
We also acknowledge the LACEGAL People Network  supported by the European Community.
Simulations were run in Fenix and HOPE clusters at IAFE and Cecar cluster at University
of Buenos Aires, Argentina.
\bibliographystyle{mn2efix.bst}

\bibliography{references}

\section*{Appendix: A parametric toy model constrained by observations}

\begin{figure*}
\vspace{6.5cm}
\includegraphics{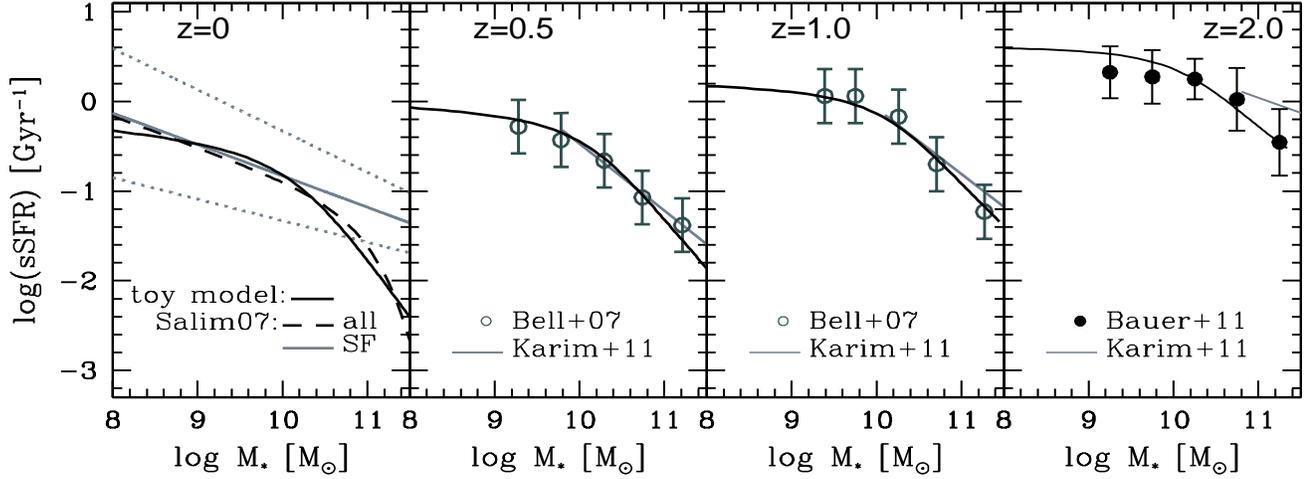}
\caption[]
{Average sSFR vs \ms\ for all (star forming and passive) observed galaxies at four different 
epochs (lines and symbols, the references are given inside the panels). At 
$z=0$, the fit given in \citet{Salim+2007} for star-forming galaxies is also shown 
with a grey line (dotted lines depict the standard deviation).  As can be appreciated, at low masses, most of the galaxies are
actually star-forming. The solid curves correspond to the isochrones of our parametric evolutionary
toy model, constrained just to fit the empirical sSFR--\ms\ and \ms--\mh\ relations at different
epochs. For the fits to the latter relation, see Fig. \ref{fig:fivsMh}. 
}
\label{Mod-sSFR}
\end{figure*}

In the following, we give a brief description of the parametric toy model of galaxy mass growth constrained
by the sSFR(\ms,$z$) and \ms(\mh,$z$) empirical relations. This model was used for comparisons
with our simulations in Sec. 3 and 4.

We start by assuming that the baryonic infall rate is initially driven by the dark matter halo aggregation rate: 
$\dot{M}_{b}(z)= F_{b,U} \times \dot{M}_{\rm vir}(z)$,
where $F_{b,U} = 0.15$ is the universal baryonic fraction, and $\dot{M}_{\rm vir}(z)$ is the average halo
mass aggregation rate.  The fits to the results of the Millennium Simulation, 
given in \citet{Fakhouri+2010} as a function of mass and $z$, are used here for deriving $\dot{M}_{\rm vir}(z)$. 
We define the galaxy SFR as a function of \mh\ and $z$ as:
\begin{equation}
SFR(\mh, z) \equiv \dot{M}_{b}\times \mathcal{T}(\mh,z)\times\epsilon(z),
\end{equation}
where $\mathcal{T}(\mh, z)$ is associated to the shape of the SF efficiency function and $\epsilon(z)$ gives its normalization at each epoch. 
The SF efficiency function encodes all the highly complex astrophysical mechanisms that affect the stellar mass assembly in galaxies. We 
assume that it affects in a different way galaxies of different masses: for low-mass galaxies, the UV background and the 
stellar-driven feedback produce a reduction of the SF; for high masses, the long cooling times for the gas
and the strong AGN-driven feedback diminish the SF in galaxies, too. 
Therefore, we assume that $\mathcal{T}(\mh, z)$ is a double power (bell-shaped) function:
\begin{equation}
\mathcal{T}(\mh, z) = \frac{2\mathcal{T}_{0}}{\left[\left(\frac{\mh}{M_{1}} \right)^{-\alpha}+\left(\frac{\mh}{M_{1}} \right)^{\beta} \right]},
\end{equation}
normalized in such a way that $\mathcal{T}(M_{\rm max}, z)=1$, where $M_{\rm max}$ is the mass at which the function 
has its maximum, and is related to the other parameters by $M_{\rm max}=(\alpha/\beta)^{1/(\alpha+\beta)}M_1$. 
Therefore, $\mathcal{T}$ has 3 independent parameters once normalized to 1 at the maximum.
We allow the parameters $\alpha$, $\beta$, and $M_{\rm max}$ to change with time as linear
functions of log($1+z$), ($1+z$), and $\propto c_1 + c_2(1+z)^{\gamma}$, respectively.

In the following, we assume that \ms\ actually grows by in-situ and
ex-situ modes; the latter represents the accretion of stars formed outside (mergers):
\begin{equation}
\Delta \ms = SFR\times\Delta t\times(1-R) + \Delta M_{\rm ex}(\mh,z),
\end{equation}   
where $R=0.45$ is the average recycling factor due to stellar mass loss. We assume that SFR is constant over a period of 
$\Delta t = 0.1$ Gyr. The ex-situ mode is parametrized by a function such that it increases
with \mh\ and time since $z\sim 2$. 
The functionalities we have introduced for the evolution of $\alpha$, $\beta$, $M_{\rm max}$, and $M_{\rm ex}(\mh,z)$
are based on very general theoretical and observational arguments.  
All the parameters associated to the SF efficiency and ex-situ growth functions are constrained to 
reproduce both the sSFR(\ms) and \ms(\mh) empirical relations at different epochs (isochrones).  
A detailed analysis of the constrained SF efficiency as a function
of \mh\ and $z$ is presented in Gonz\'alez-Samaniego et al. (in prep.).  
An important feature of this function is that the highest efficiencies
at all epochs (since $z=4$) correspond to halos of masses $\approx 0.8-1\times 10^{12}$ \msun. 
For larger masses, the efficiency
decreases with mass, more and more as $z\rightarrow 0$; for much smaller masses, the efficiency decreases for lower
masses, more and more as $z$ increases.

\begin{figure} 
\vspace{7.cm}
\includegraphics{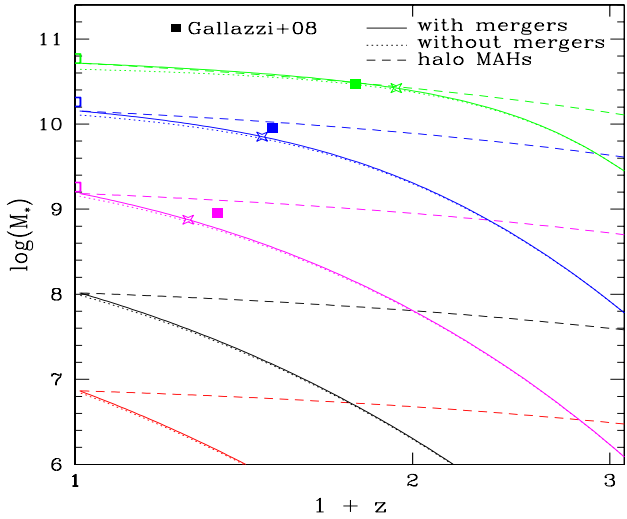}
\caption{
Average stellar mass assembly tracks inferred by means of our evolutionary parametric
toy model constrained by observations (solid curves). Dotted curves show the component
of these tracks due to in-situ SF. The dashed curves depict the corresponding halo MAHs
renormalized in each case in order that \mh($z=0$) = \ms($z=0$).  For smaller masses,
the delay of the stellar mass growth with respect to the halo growth increases.  Open stars correspond to 
the $z$ at which half of the \ms\ was assembled. This characteristic $z$ estimated for SDSS galaxies by means
of the archeological approach \citep{Gallazzi+2008} is shown with solid squares
for three samples of galaxies, each one with an average present-day \ms\ indicated
by the open squares.
}
\label{Mod-Msz}
\end{figure}

In Figs. \ref{Mod-sSFR} and \ref{fig:fivsMh}, we plot the sSFR vs \ms\ and \fs\ vs \mh\
isochrones obtained with this model at four epochs (blue dashed curves), respectively. 
The observational data related to the
{\it total} sSFR -\ms\ relation that we have used are taken from:
\citet{Salim+2007} for SDSS/GALEX galaxies at $z\approx 0$; 
\citet{Bell+2007} for COMBO-17 galaxies at redshift bins centred at $z\sim 0.5$ and $z\sim 1$; 
\citet{Karim+2011} for galaxies in the
COSMOS field (instead of averages in mass bins, a 1.4 GHz (VLA) image stacking analysis was performed; 
in this case, we plot the fits to the data given in the paper); 
and \citet{Bauer+2011} for a stellar mass-selected sample of galaxies from the GOODS NICMOS survey 
(their $1.5<z<2$ redshift bin is used; the data in their next bin, $2.0<z<2.5$, are actually very similar). 
We have converted all the samples to a \citet{Chabrier2003} IMF. 
In the literature, there are many other determinations of the sSFR-\ms\ relation at different epochs (see
the Introduction for references) but most of them refer only to star-forming galaxies. Actually, at masses below a few
$10^{10}$ \ms, the great majority of galaxies are blue/star forming, while at larger masses, the majority are red/passive
galaxies. 
At $z>3$, even these massive galaxies can be blue/star forming. Therefore, 
we take care that our toy model
agrees with the few available observational estimates of the sSFR-\ms\ relation at high $z$ (see Fig. \ref{fig:sSFR-s320}). 

Regarding the empirical \fs--\mh\ relations, we have used the continuous function
(red triple-dot-dashed curves in Fig. \ref{fig:fivsMh}) proposed in FA10 at $z=0-4$ 
for describing the abundance matching results of
\citet{Behroozi+2010} at two different redshift ranges: $0<z\lesssim 1$ and $1\lesssim z<4$. At the low-mass side, 
the decrease with $z$ of the \fs(\mh,$z$) function used in FA10 is faster than the one obtained by Behroozi et al. (2010, 2012) 
at $0<z\lesssim 1$, but  slower than what is reported in \citet{Yang+2012}. 

In Fig. \ref{Mod-Msz}, five stellar mass growth histories obtained with this model are plotted. 
Solid lines depict the total (in-situ + ex-situ) mass growths, 
while dotted lines correspond only to the in-situ SF mode. 
As is seen, the ex-situ (dry merger) mode is moderately significant
only for the most massive galaxy models. For present-day masses of $\ms\approx 5\times 10^{10}$ \msun, mergers contributed 
with the $\approx 17\%$ to the total stellar mass assembly since $z=2$. Along each track, we plot also the $z$ where half of 
the present-day \ms\ has been attained (open stars).
This characteristic redshift has been also inferred by means of stellar population synthesis models (archaeological approach) 
and determined for a large sample of SDSS galaxies in \citet{Gallazzi+2008}. 
Their results for three average present-day masses (open squares),
close to our three most massive models, are represented in Fig. \ref{Mod-Msz} with solid squares. The agreement of this
quantity between our models and the archaeological inferences is encouraging. We also plot for each of our model tracks their
corresponding mean halo MAHs (dashed lines) but shifted down vertically in each case in order that $\mh($z=0$) = \ms($z=0$)$.
In this way, it is possible to compare the shapes of the \ms\ and \mh\ growth histories for each model. As the mass decreases, 
the delay of the active phase of stellar mass growth with respect to the halo one is more significant. The four 
MAHs for the
lower stellar masses in Fig. \ref{Mod-Msz}, as well as the corresponding histories associated to \ms/\mh, were reproduced in Figs. \ref{fig:massagg}, 
\ref{fig:massaggratio_v2}, and \ref{fig:Fratio}, in order to compare them with the results obtained in the simulations (Sec. \ref{comparison}). 


\end {document}